\documentclass{article}

\usepackage[a4paper, margin=3.5cm]{geometry}

\usepackage{orcidlink,thumbpdf}
\usepackage{amsmath, amssymb}
\usepackage{array}
\usepackage{graphicx}
\usepackage{natbib}
\usepackage{hyperref}
\usepackage{lmodern}

\usepackage{xcolor} % Load the xcolor package
\usepackage{xparse} % Load the xparse package
\usepackage{ulem}
\usepackage{tablefootnote}
\usepackage{listings}

\definecolor{matlabBlue}{rgb}{0.0,0.0,0.6}

\definecolor{matlabGreen}{rgb}{0.0,0.5,0.0}

\definecolor{matlabPurple}{rgb}{0.58,0.0,0.82}

\definecolor{matlabGray}{rgb}{0.5,0.5,0.5}

\definecolor{matlabBlack}{rgb}{0,0,0}

\definecolor{matlabBackground}{rgb}{0.97,0.97,0.97}

\lstdefinestyle{matlab}{
  language=Matlab,
  basicstyle=\ttfamily,
  basewidth=0.5em,
  escapeinside={(*@}{@*)},
  alsoletter={.},
  backgroundcolor=\color{matlabBackground},
  keywordstyle   =\color{matlabBlack},
  keywordstyle   =[2]\color{matlabPurple}\bfseries,
  morekeywords=[2]{citiesItaly2024.mat},
  commentstyle=\color{matlabGreen}\itshape,
  stringstyle=\color{matlabPurple},
  numberstyle=\tiny\color{matlabGray},
  numbers=none,
  stepnumber=1,
  numbersep=8pt,
  showstringspaces=false,
  breaklines=true,
  frame=single,
  rulecolor=\color{black},
  tabsize=2,
  captionpos=b,
  xleftmargin=2em,
  framexleftmargin=1.5em
}

\newcommand{\mcode}[1]{\lstinline[style=matlab]!#1!}
\newcolumntype{L}[1]{>{\raggedright\arraybackslash}p{#1}}
\usepackage{framed}

\newcommand{\pkg}[1]{{\normalfont\fontseries{b}\selectfont #1}}
\let\proglang\textsf

\newcommand{\Keywords}[1]{\par\noindent\textbf{Keywords:} #1}

\shortcites{Robook:2024, greenacre2022principal, Riani2025DataScienceWithMATLAB,
yohai:2019, RCPT:2015, h+r+v:2005, tool:2012, riani2015forward,
APTarxiv:2023, arc:2004, tenenbaum2000global, zeileis2020colorspace,
pedregosa2011scikit, factoextra, factoshiny, rrcov, mathworksPCA}

\author{
Marco Riani\orcidlink{0000-0001-7886-2207}\\
Universit\`a di Parma
\and
Anthony C. Atkinson\orcidlink{0000-0001-7937-9225}\\
London School of Economics
\and
Aldo Corbellini\orcidlink{0000-0002-6936-7295}\\
Universit\`a di Parma
\and
Domenico Perrotta\orcidlink{0000-0002-0987-3823}\\
EC Joint Research Centre (JRC)
\and
Francesca Torti\orcidlink{0000-0003-0801-1735}\\
EC Joint Research Centre (JRC)
\and
Gianluca Morelli\orcidlink{0000-0003-1297-5385}\\
Universit\`a di Parma
}
\title{Interactive and Robust PCA Graphics with FSDA in \proglang{MATLAB}}

\date{}

\begin{document}

\maketitle

\begin{abstract}
Principal Component Analysis (PCA) is a standard method for dimension reduction and the exploratory analysis of multivariate data, but a complete applied workflow often requires more than numerical eigenvectors and static plots. This article documents a set of routines in the \pkg{FSDA} toolbox for \proglang{MATLAB} that implements an integrated workflow for interactive and robust PCA graphics. The software contribution consists of coordinating classical SVD-based PCA, dynamic biplot scaling controlled by row and column parameters, score-distance and orthogonal-distance diagnostics, hard-trimming robust PCA through the Minimum Covariance Determinant and the Forward Search, linked brushing, geographical displays, and three-dimensional projection diagnostics. The implementation returns reproducible numerical output while also exposing graphical controls for changing biplot geometry, confidence regions, displayed labels, robust trimming levels, and linked selections. The methodology is illustrated on quality-of-life indicators for Italian provinces, where the software reveals variable groups, atypical provinces, robust and non-robust PCA structures, and geographical patterns. The emphasis is on statistical software: no new PCA estimator is proposed, but established PCA and robust-PCA methodology is combined in a documented, extensible, and script-reproducible \proglang{MATLAB} environment.
\end{abstract}


%% -- Introduction -------------------------------------------------------------

%% - In principle "as usual".
%% - But should typically have some discussion of both _software_ and _methods_.
%% - Use \proglang{}, \pkg{}, and \code{} markup throughout the manuscript.
%% - If such markup is in (sub)section titles, a plain text version has to be
%%   added as well.
%% - All software mentioned should be properly \cite-d.
%% - All abbreviations should be introduced.
%% - Unless the expansions of abbreviations are proper names (like "Journal
%%   of Statistical Software" above) they should be in sentence case (like
%%   "generalized linear models" below).

\section{Introduction}
\label{introsec}

Principal Component Analysis (PCA) is a powerful and versatile statistical method for the dimension reduction of two-way data tables when the rows are units and the columns are variables. \citet{greenacre2022principal} provide a wide-ranging review of PCA and related methods, including correspondence analysis and extensions to images, shapes, and functions. Their review also illustrates the broad software support for PCA, especially in \proglang{R} and \proglang{Python}. The present paper complements that perspective by documenting a \proglang{MATLAB} implementation in which PCA computation, robust diagnostics, interactivity, and linked graphics are handled in one statistical software workflow.

PCA is already available in many environments. The standard \proglang{MATLAB} Statistics and Machine Learning Toolbox provides PCA computation and basic plots \citep{mathworksPCA}, and specialized \proglang{MATLAB} tools for exploratory multivariate analysis have also been proposed \citep{ballabio:2015}. In \proglang{R}, packages such as \pkg{ade4} \citep{dray2007ade4}, \pkg{FactoMineR} \citep{factominer2008}, and \pkg{factoextra} \citep{factoextra} provide extensive multivariate analysis and graphical summaries, while \pkg{Factoshiny} offers a \pkg{shiny} interface to selected \pkg{FactoMineR} analyses \citep{factoshiny}. Domain-specific interactive PCA is available, for example, in the Bioconductor package \pkg{pcaExplorer} for RNA-seq data \citep{pcaexplorer2019}. Robust PCA is implemented in packages such as \pkg{rrcov}, which includes high-breakdown estimators and the ROBPCA approach of \citet{h+r+v:2005} \citep{rrcov}. In \proglang{Python}, \pkg{scikit-learn} provides PCA estimation, with interactivity usually supplied by separate plotting libraries or notebooks \citep{pedregosa2011scikit}.

Table~\ref{tab:software_overview} summarizes this software context. The novelty of the implementation described in our paper is not the proposal of a new PCA estimator. Instead, it is the integration, within the Flexible Statistics and Data Analysis (\pkg{FSDA}) toolbox for \proglang{MATLAB} (Section \ref{FSDAsec}), of dynamic biplot geometry, robust hard-trimming diagnostics, Forward Search monitoring, linked brushing, and optional geographical displays in a single \proglang{MATLAB} workflow.

\begin{table}[t!]
\centering
\caption{Position of the \pkg{FSDA} contribution within existing PCA software. The table is intentionally comparative rather than exhaustive.}
\label{tab:software_overview}
\small
\vspace{3mm}
\resizebox{\textwidth}{!}{%
\begin{tabular}{|L{3.5cm}|L{3.5cm}|L{3.5cm}|L{3.5cm}|}
\hline
\textbf{Software context} & \textbf{PCA and graphics} & \textbf{Robustness and interactivity} & \textbf{Present contribution} \\
\hline
\proglang{MATLAB} standard tools and related toolboxes & PCA, score plots, loading plots, biplots, and exploratory routines. & Interactivity is mainly supplied by the general graphics system; robust PCA and linked diagnostics are not a single workflow. & \pkg{FSDA} adds dynamic biplot controls, robust hard trimming, Forward Search monitoring, and linked graphics. \\
\hline
\proglang{R}: \pkg{ade4}, \pkg{FactoMineR}, \pkg{factoextra}, \pkg{Factoshiny} & Extensive PCA, correspondence analysis, factorial methods, and static or \pkg{shiny}-based graphical summaries. & Strong visualization support; robust PCA and Forward Search monitoring are not the central design target. & \pkg{FSDA} complements these tools by focusing on robust interactive PCA in \proglang{MATLAB}. \\
\hline
\proglang{R}: Bioconductor \pkg{pcaExplorer} & Interactive PCA exploration for RNA-seq analyses. & Rich domain-specific interaction and reporting. & \pkg{FSDA} is general multivariate software rather than application-specific. \\
\hline
\proglang{R}: Robust PCA software such as \pkg{rrcov} & Robust covariance and robust PCA output, including ROBPCA-type diagnostics. & High-breakdown estimation is central; no interactivity nor dynamic plots. & \pkg{FSDA} combines robust estimation with dynamic biplots and linked views. \\
\hline
\proglang{Python}: \pkg{scikit-learn} plus plotting libraries & Efficient PCA estimation and general plotting in scripts or notebooks. & Interactive graphics typically require external libraries and user assembly. & \pkg{FSDA} provides a packaged PCA graphics workflow inside \proglang{MATLAB}. \\
\hline
\pkg{FSDA} PCA tools documented here & Classical PCA output, ``smart'' eigenvector plots, outlier maps, dynamic biplots, and three-dimensional projection plots. & MCD, Forward Search, linked brushing, scatter plot matrices, geomaps, geobubbles, and reproducible scripts. & This paper documents the methodology, implementation, and applied use of this integrated functionality. \\
\hline
\end{tabular}
}
\end{table}

The paper has four specific contributions. First, it specifies the PCA methodology implemented in the new \pkg{FSDA} functionality: classical PCA, biplot geometry, outlier maps, robust PCA by hard trimming, and linked visual diagnostics. Secondly, it describes how the functionality is implemented, including the main functions, returned objects, graphical controls, and extensibility conventions. Thirdly, it demonstrates how the software is used in practice on real data, with code that moves from pairwise exploration to classical PCA, robust PCA, brushing, geography, and three-dimensional diagnostics.
Finally, it explains how the replication material has been organized around a
single commented \proglang{MATLAB} script, \texttt{replication.m}, which is the
primary entry point for reproducing the computational workflow and launching the
interactive displays used in the paper.

The rest of the paper is organized as follows. Section~\ref{PCAsec} reviews the PCA geometry needed for the implementation. Section~\ref{methodssec} describes the methodology implemented in the software, and Section~\ref{FSDAsec} describes the implementation design and user-facing functionality. Sections~\ref{datasec}--\ref{3sec} present the case study and show how the tools are used in practice. Section~\ref{sec:replication} summarizes the replication material. Section~\ref{futuresec} concludes with limitations and possible extensions.

\section{Principal Component Analysis and Graphical Representations}
\label{PCAsec}

Given $p$ quantitative variables $X_1, X_2, \ldots, X_p$, the goal of Principal Component Analysis (PCA) is to replace the original, possibly correlated, variables with a new set of variables, the principal components, which satisfy \citep{jol:2002}
\begin{itemize}
  \item they are uncorrelated (orthogonal);
  \item they are ordered by decreasing explained variance.
\end{itemize}

 The first principal component is the linear combination of the $p$ variables having maximal variance.  The second principal component is the linear combination with the next largest variance and is uncorrelated with the first  and so on up to the $p$-th component.  If the original $p$ variables are highly correlated, then the first $q<p$ components capture most of the total variance, so that these $q$ components provide a reduced-dimensional approximation to the data structure. The primary purpose of this technique is to reduce a large number of variables, each representing a feature of the phenomenon, into a few latent variables.    Dimensionality is reduced by retaining only the leading (by variance) new variables.  Here ``explained variance'' is more formally the sum of the squared distances of the points from their multidimensional mean.

For the procedure to provide a useful analysis, the variables $X$ need scaling to have zero mean. Unless the variables are measurements of similar quantities, the variables should also be scaled to have unit variance.
From the point of view of robustness and outlier detection these scalings will be distorted by outliers which will affect both the estimated means and variances.  We explore this point in \S\ref{robsec}. \citet{h+r+v:2005} provide a description of the problem and one robust solution.

PCA can be applied to  $\tilde X$ (the matrix of deviations from the mean) or $Z$ (the matrix of standardized deviations), both  matrices of dimension $n \times p$. For the standardized deviations $Z$,
the principal components are obtained as linear combinations of the standardized variables, with weights given by the elements of the eigenvectors of $Z^\top Z/(n-1) = R$, where $R$ is the covariance matrix of $Z$ (or the correlation matrix of $\tilde X$).
The corresponding eigenvalues are proportional to the amount of variance explained by each principal component. The sum of the eigenvalues equals the total variance, which is equal to the sum of the variances (i.e., the trace) of the correlation matrix and to the sum of the squared distances of the observations from their multidimensional mean.

% If, as is often the case, the first two principal components explain an appreciable part of the variance of the data, for example 70\%,  the data, both responses and variables, can usefully be plotted in the space of the first two principal components.  The ``loading plot'' is a graphical representation of the  correlations between the original variables and the first two principal components via vectors (arrows).  The data are represented, customarily by circles, in a score plot, where the scores are the coordinates of the original observations in the  principal component space.
%

If, as is often the case, the first two principal components explain an appreciable part of the variance of the data, for example 70\%, the data (both responses and variables) can be represented in the space spanned by these components. A ``loading plot'' provides a graphical representation of the correlations between the original variables and the selected principal components,  through vectors (arrows) whose directions and lengths reflect their contributions. The observations themselves can be represented, customarily by circles, through their coordinates (the scores) in the same reduced principal component space.
When both representations are displayed jointly, the resulting display is commonly referred to as a ``biplot'', which allows the simultaneous exploration of relationships among variables, among observations, and between the two. This joint representation plays a central role in exploratory PCA and provides a natural framework for the interactive visual tools discussed in the following sections.

\section{Methodology implemented in FSDA}
\label{methodssec}

This section identifies the statistical methodology implemented in the new \pkg{FSDA} PCA functionality. The purpose is not to introduce a new PCA estimator, but to document the methods that have been brought together in the software and exposed through reproducible output and interactive graphics.

\subsection{Classical PCA output}

The function \mcode{pcaFS} computes a classical PCA from a numeric matrix or a table. The analysis can be based on centered variables or on standardized variables, depending on the scale of the input and the options used. The main numerical objects are the correlation or covariance matrix, eigenvalues, explained variance, eigenvectors, loadings, communalities, scores, score distances, and orthogonal distances. These quantities are printed in the \proglang{MATLAB} Command Window and are also returned in a structured object, so that interactive exploration is accompanied by script-level reproducibility. Appendix~\ref{sec:command_window} shows a typical output structure.

\subsection{Dynamic biplot geometry}

The interactive biplot is based on the singular value decomposition described in Section~\ref{PCAsec}. The software implements a family of biplots indexed by two parameters, $\alpha$ and $\omega$, both in $[0,1]$. The parameter $\alpha$ controls how the singular values are distributed between row points and variable arrows, and hence changes the emphasis between observation distances and variable interpretation. The parameter $\omega$ redistributes the global factor $\sqrt{n-1}$ between row and column coordinates, and therefore changes the visual balance of the joint display. The function \mcode{biplotFS} exposes these choices through sliders, while \mcode{pcaFS} launches the chosen biplot as part of a complete PCA analysis.

\subsection{Outlier diagnostics and robust PCA}

The implemented diagnostic map follows the score-distance/orthogonal-distance decomposition of \citet{h+r+v:2005}. The score distance measures distance within the retained principal-component subspace, while the orthogonal distance measures lack of fit to that subspace. This separation is important because an observation can be extreme in the PCA plane, poorly represented by the plane, or both.

Robust PCA is implemented through hard trimming. In the MCD branch, PCA is computed from the subset of observations retained by the Minimum Covariance Determinant estimator for a user-specified breakdown point. In the Forward Search branch, PCA is recomputed along an increasing sequence of subset sizes. The Forward Search is therefore a monitoring device: it shows how eigenvalues, scores, distances, and the biplot change as units enter or leave the subset. The two robust branches answer complementary questions. MCD gives a fixed high-breakdown fit, while the Forward Search displays a path of fits indexed by the amount of trimming.

\subsection{Linked views, brushing, and geographical information}

The implemented graphical methodology treats interactivity as part of the statistical analysis rather than as a final decoration. A selected set of row indices can be propagated from the biplot to a scatter plot matrix, to robust diagnostic panels, and, when geographical information is supplied, to geomaps or geobubble plots. Conversely, the same row labels and selected units remain available for inspection in the returned output. In this way brushing becomes a reproducible device for defining subsets whose statistical and geographical properties can be examined in linked displays.

\section{Software design, functionality, and extensibility}
\label{FSDAsec}

% DP: first paragraph changed because FSDA was already introduced in the Introduction and mentioned several time earlier. 

%All computations and graphical analyses presented in this paper are carried out using the Flexible Statistics and Data Analysis (\pkg{FSDA}) toolbox for \proglang{MATLAB}. \pkg{FSDA} is an open-source collection of functions for robust and exploratory data analysis, with a particular focus on diagnostic tools, interactive graphics, and reproducible workflows \citep{tool:2012}. 

\pkg{FSDA}, which is used for all computations and graphical analyses presented in this paper, is an open-source collection of \proglang{MATLAB} functions for robust and exploratory data analysis, with a particular focus on diagnostic tools, interactive graphics, and reproducible workflows \citep{tool:2012}. 
The toolbox includes implementations of classical and robust methods for regression, multivariate analysis, clustering, and dimension reduction. Related computational and methodological developments are described in \citet{riani2015forward}, \citet{RCPT:2015}, \citet{APTarxiv:2023}, and \citet{Riani2025DataScienceWithMATLAB}.

The PCA functionality described here follows a layered design. First, the input layer accepts a numerical matrix or a \proglang{MATLAB} table, with optional row labels, variable labels, and geographical information. Second, the numerical layer computes PCA quantities, robust subsets, distances, and monitoring output. Third, a common output structure stores the numerical results in array and table form. Fourth, the graphical layer builds figures, GUI controls, and callbacks from the stored results. This separation is important because the same analysis can be run as a script for replication or opened interactively for exploration.

\begin{table}[t!]
\centering
\caption{Main \pkg{FSDA} functions and capabilities used in the paper. The table identifies the role of each component in the implemented workflow.}
\label{tab:functionality_main}
\small
\resizebox{\textwidth}{!}{%
\begin{tabular}{|L{3.5cm}|L{3.5cm}|L{3.5cm}|L{3.5 cm}|}
\hline
\textbf{Function or capability} & \textbf{Role in the workflow} & \textbf{Main returned or displayed information} & \textbf{Interactive or extensible features} \\
\hline
\mcode{spmplot} & Pairwise exploration before PCA. & Scatter plot matrix, correlations, variable ordering, and optional superimposition of ellipses or contours, backgrounds depending on correlation. & Alternative upper/lower-panel displays, AOE ordering (\S~\ref{2wIsestrucsec}), linked highlighting. \\
\hline
\mcode{pcaFS} & Main PCA entry point. & Explained variance, eigenvectors, loadings, communalities, scores, distances, outlier map, and biplot. & Launches ``smart'' eigenvector plots, dynamic biplot, robust panels, maps, and brushing. \\
\hline
\mcode{biplotFS} & Standalone biplot engine. & Row coordinates, arrow coordinates, labels, confidence ellipse, and figure handles. & Sliders for $\alpha$ and $\omega$, sign changes, label density, coloring by diagnostics, brushing. \\
\hline
MCD panel in \mcode{pcaFS} & Fixed robust PCA fit. & Robust subset, robust distances, outlier flags, and robust biplot. & Breakdown-point and confidence-level controls; robust/non-robust distance comparison. \\
\hline
Forward Search panel in \mcode{pcaFS} & Monitoring robust PCA along subset sizes. & Evolution of distances and PCA structure as $m$ increases. & Subset-size slider, automatic mode, and monitoring plots linked to the biplot. \\
\hline
Geographical options & Spatial interpretation of PCA results. & Geomap or geobubble displays based on scores, selected units, and shapes or coordinates. & Linked brushing between PCA plots, scatter plot matrices, and maps. \\
\hline
\mcode{pcaProjection} & Three-dimensional projection diagnostics. & Original and projected points, fitted subspaces, and confidence ellipsoids. & Rotatable three-dimensional displays and assessment beyond two PCs. \\
\hline
\end{tabular}
}
\end{table}

The returned structure is the main interface between numerical and graphical components. For example, \mcode{out.score}, \mcode{out.loadings}, \mcode{out.orthDist}, and \mcode{out.scoreDist} can be used directly in scripts, while the associated graphical callbacks use the same quantities to update labels, colors, confidence regions, and brushed selections. This design makes it possible to reproduce the figures non-interactively and then to open the same objects for interactive exploration.

Extensibility is supported by using common data structures for centers, loadings, scores, distances, subset membership, and outlier indicators. New robust estimators can be connected to the existing graphics provided that they return these quantities in the same form. Similarly, new linked displays can use the same vector of selected row indices as the biplot, scatter plot matrix, and geographical displays.

The graphical defaults also follow a color-use policy. Qualitative differences, such as brushed and unbrushed observations, should be encoded not only by hue but also by marker type or fill. Ordered magnitudes, such as orthogonal distances, should use perceptually ordered sequential palettes. Signed quantities, such as correlations or centered scores, should use diverging palettes with a meaningful center.
This policy follows the recommendations and diagnostic tools of \citet{zeileis2020colorspace}, whose palettes are built by interpolating in the perceptually based hue--chroma--luminance (HCL) color space rather than directly in RGB. As a consequence, equal steps in the data correspond to approximately equal perceived color differences. This approach is  consistent with broader recommendations on scientific color maps \citep{Zeileis2009,Crameri2020}. \label{page:HCL}
%This policy follows the HCL-based recommendations and diagnostic tools of \citet{zeileis2020colorspace}, and is also consistent with broader recommendations on scientific color maps \citep{Zeileis2009,Crameri2020}.
The replication scripts accompanying this paper are intended to \textit{optionally} regenerate manuscript figures with these palette choices, controlled through a single \texttt{colorBlindSafe} argument exposed by the relevant \pkg{FSDA} functions (e.g., \texttt{spmplot}, \texttt{pcaFS}). For backward compatibility, setting \texttt{colorBlindSafe} to \texttt{false} reproduces the legacy palettes used in earlier releases of \pkg{FSDA}.

To make the enlarged set of options easier to use, the implementation also
includes a \proglang{MATLAB} JSON signature file specifying the admissible
name-value arguments and, when appropriate, their possible values. This file is
used by the \proglang{MATLAB} editor to provide automatic completion when the
user types a call to functions such as \mcode{pcaFS}. Thus, after the required
input argument has been entered, the editor suggests the available options; once
an option has been selected, it also suggests the admissible values or value
types. This improves discoverability of the interface, reduces the need to
consult the help file for routine usage, and helps keep the syntax of the
interactive PCA tools consistent across the FSDA functions. An example is shown in Appendix Figure~\ref{fig:jsoncompletion}.

\section{Case study: Quality of Life in Italian Provinces}
\label{datasec}

The case study demonstrates how the new \pkg{FSDA} functionality is used in practice. The data come from the annual ``Quality of Life'' (QoL) survey of the Italian financial newspaper ``Il Sole 24 ORE'', which ranks the 107 Italian provinces using 90 indicators of personal and family well-being.
 The newspaper ranking rescales each indicator to the interval $[1,1000]$, changes signs where necessary so that larger values correspond to better quality of life, and averages the resulting scores. Such a ranking is easy to communicate, but it can hide correlation structure and can be sensitive to outlying values. PCA provides an alternative exploratory view because it summarizes common variation among indicators, while the robust and interactive tools in \pkg{FSDA} allow the analyst to check whether the summary is stable.

The analysis is organized as a reproducible workflow addressing five questions: (i) whether the indicators have a low-dimensional correlation structure; (ii) which variables define the dominant components; (iii) which provinces are atypical in score space or in orthogonal distance from the PCA subspace; (iv) whether the conclusions are stable under robust hard trimming; and (v) how the PCA structure is related to geography. The GitHub  repository
\url{https://github.com/UniprJRC/PCAinteractive} contains the
\proglang{MATLAB} source files used to generate the figures in the paper (more details are in \S
\ref{sec:replication}).

We selected 12 of the indicators listed by the newspaper from the 2024 edition of the survey, chosen subjectively to provide a balanced subset of the original indicators. The variables are presented in Table~\ref{table:QoL}.

\begin{table}[tb!]
\begin{center}
\resizebox{\textwidth}{!}{%
\begin{tabular}{p{0.72\linewidth} p{0.22\linewidth}}
\hline
\textbf{Description} & \textbf{Variable name} \\
\hline
Bank Deposits & Deposit \\
Bankrupt Companies & Bankrup \\
Index of Urban Fragility & UrbanFra \\
Invoice Payments within 30 Days & Paym30D \\
Electoral Participation & ElecPar \\
Quality of Life of Children, Young People and the Elderly & QualLif \\
Protests per capita\tablefootnote{A protest of non payment is a formal, usually notarized, declaration that a negotiable instrument, such as a cheque or promissory note, has been presented for payment and subsequently dishonored. It serves as official evidence that payment was demanded and refused. Not to be confused with street protests.} & Protest \\
Average Annual Salary & SalaryA \\
Family Spending & SpendingA \\
Employment Rate & Employm \\
Value Added per capita & AddedVa \\
Families with low ISEE\tablefootnote{Indicatore della Situazione Economica Equivalente (Indicator of Equivalent Economic Situation). It is an official metric used to evaluate the economic and asset situation of a household in order to determine eligibility for discounted social benefits and public services.} & LowISEE \\
\hline
\end{tabular}
}
\end{center}
\caption{Quality of Life (QoL) variables, source: Il Sole 24 ORE.}
\label{table:QoL}
\end{table}

When the names of the provinces are included, the data matrix is $107\times 12$. The objective of the analysis is not to produce a definitive socio-economic ranking, but to demonstrate how the implemented software moves from numerical PCA output to robust and linked graphical interpretation.

\subsection{Step 1: pairwise structure and variable ordering}
\label{2wIsestrucsec}

\begin{figure}[tb!]
\centering
\includegraphics[width=1\textwidth,clip=true]{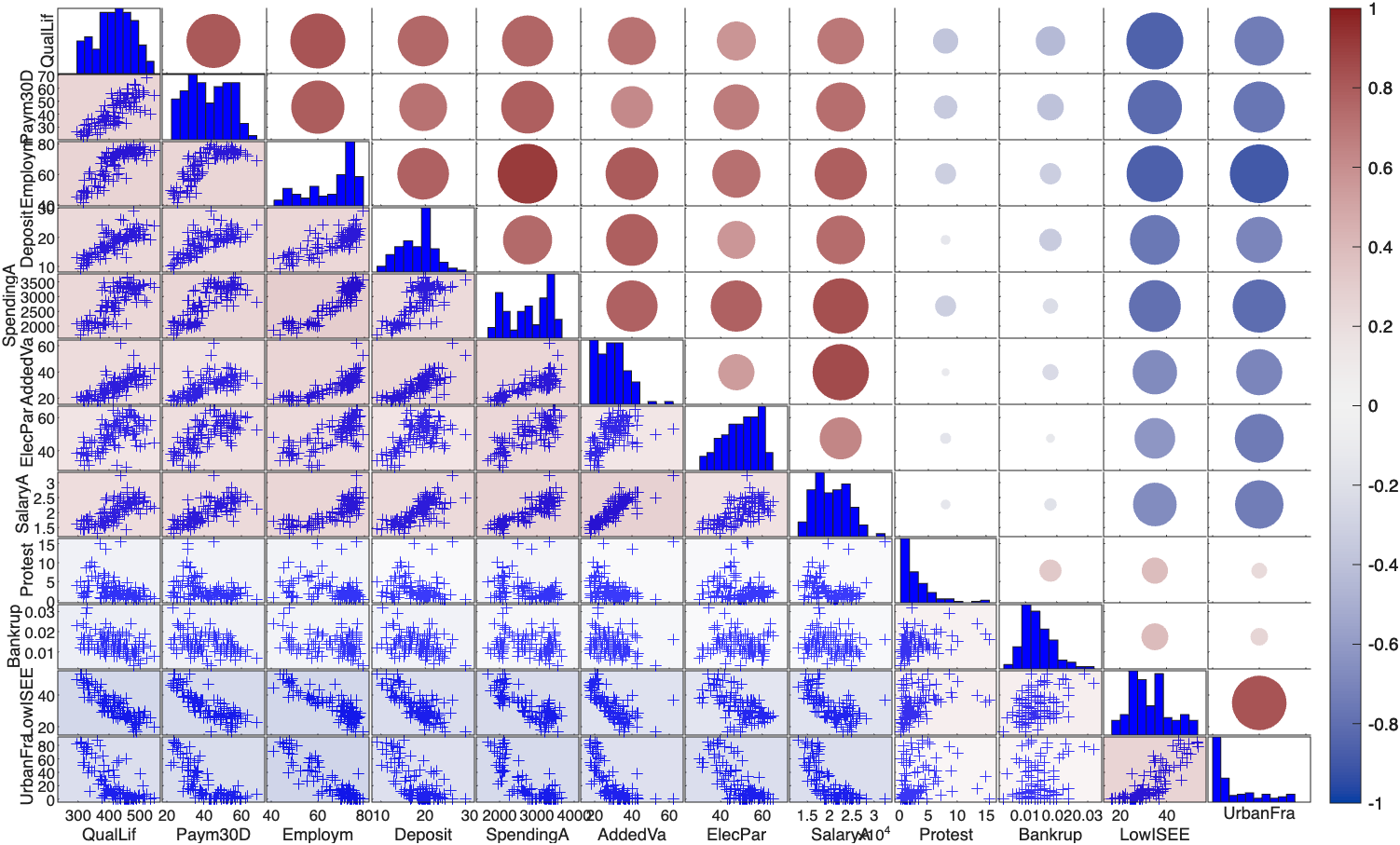}
\caption{Modified scatter plot matrix with reordered columns. Option \mcode{'colorBlindSafe'} is set to \texttt{true}.}
\label{spmAOE}
\end{figure}

We begin by examining pairwise relationships among variables to assess the overall correlation structure, and for this we look at the scatter plot matrix. That for the variables in the order given above is shown as Figure~\ref{A1} (in  Appendix A). The code to produce this Figure uses FSDA function \mcode{spmplot}:
\begin{lstlisting}
load citiesItaly2024.mat
spmplot(citiesItaly2024);
\end{lstlisting}
The cells of this plot show that many of the variables have appreciable positive or negative  correlations with each other, so that a PCA analysis can be expected to be informative.
\citet{friendly:2002:corrgrams} describes many ways in which the straightforward plot of Figure~\ref{A1} can be improved. One class of improvements is  to remove the redundancy of giving each scatter plot twice; we replace the upper-right triangle of the matrix by the magnitude and sign of the correlation between the pairs of variables. The second improvement comes from noticing that there are 12! orderings of the rows; some permutations of rows will yield more meaningful scatter plot matrices than others.  Friendly suggests ordering the variables by the angular orders of the eigenvectors from the two largest eigenvalues, that is the two most significant principal components. The order of the variables is calculated from the order of the angles $\alpha_i$,
\begin{equation}
\label{splitting}
\alpha_i  = \left\{
\begin{array}{ll}
\mbox{tan}^{-1}(v_{i2}/v_{i1}) & v_{i1} > 0 \\
\mbox{tan}^{-1}(v_{i2}/v_{i1}) + \pi  & \mbox{otherwise,}
\end{array}\right.
\end{equation}
where $v_1=(v_{11}, \ldots, v_{p1})^\top $ and $v_2=(v_{12}, \ldots, v_{p2})^\top $ are the $p \times 1$ eigenvectors associated with the two largest eigenvalues. This circular order is unfolded to a linear order by splitting at the largest gap between the adjacent vectors. An extension \citep{falissard1996spherical} is the use of the first three principal components to represent  the variables on the surface of a sphere.
Here we follow the procedure of \citet{friendly:2002:corrgrams}. The resulting extended scatter plot matrix, produced by the function \mcode{spmplot}, is shown in Figure~\ref{spmAOE}. The plot is obtained by specifying a set of optional arguments through the <\mcode{'Name',Value}> syntax\footnote{In MATLAB, optional arguments are specified using the <\mcode{'Name',Value}> format, where \mcode{'Name'} denotes the option name and \mcode{Value} its corresponding setting.
Note that, for clarity, we use the  angle brackets $\langle \cdot,\cdot \rangle$ notation for visual identification  of the option-value pairs in the text. The angle brackets are not part of the MATLAB syntax.
In recent MATLAB releases, options can also be specified using the equivalent \mcode{Name=Value} syntax.}.
In particular, the option <\mcode{'typespm',typespm}> controls the type of graphical elements used to represent the correlations in the scatter plot matrix.
If \mcode{typespm} is specified as a structure (\mcode{struct})\\[0.5mm]
\begin{lstlisting}[firstnumber=4]
typespm = struct;     typespm.upper = 'circle';
spmplot(citiesItaly2024,'order','AOE','typespm',typespm,...
                        'colorBackground',true,'colorBlindSafe',true);
\end{lstlisting}

\noindent the user can control which part of the scatter plot matrix (upper or lower triangular) is used to display a given type of graphical symbol. In this example, circles are used in the upper triangular part to represent pairwise correlations.
The option <\mcode{'order','AOE'}> applies the angular ordering discussed above, while <\mcode{'colorBackground',true}> specifies that the background of each panel is coloured according to the magnitude of the corresponding correlation coefficient.
Finally, \mcode{<'colorBlindSafe',true>} switches to %\cite{zeileis2020colorspace}'
HCL-based diverging palette (see page \ref{page:HCL}):
correlations are shown with a neutral white center at zero and colors of equal perceived lightness for positive and negative values of equal magnitude, applied consistently to both the panel background colors and the circle/square/number correlation encodings.

%\redmarginpar{xFra: In principle, colors of each pair of bg cell and circle should be identical}
The upper right triangle  of the matrix shows the correlations between the pairs of variables; the  diameter of the filled circles is proportional to the correlation, with the colour indicating the sign of the correlation and echoing the strength of the correlation.
% Option 'colorBackground',true enables to have a background color for each scatter that also reflects the magnitude of the correlation.
The top line corresponds to the variable ``Quality of Life of Children, Young People and the Elderly''. This excellent, informative plot shows that there are three groups of variables. The largest is the set of mutually positively correlated variables, which all contribute to positive effect on the quality of life. The two variables ``Protests per capita'' and ``Bankrupt Companies'' have little effect on the quality of life, whereas high values for the last two variables, ``the number of families with  low ISEE'' and ``Urban Fragility'', strongly indicate a poor quality of life.  Note that we do not have an overall variable ``quality of life'', i.e. this is not a regression analysis.

\subsection{Step 2: classical PCA output}

\begin{figure}[!htb]
\centering
\includegraphics[trim=0cm 10.5cm 0cm 11cm, clip, width=1\textwidth]{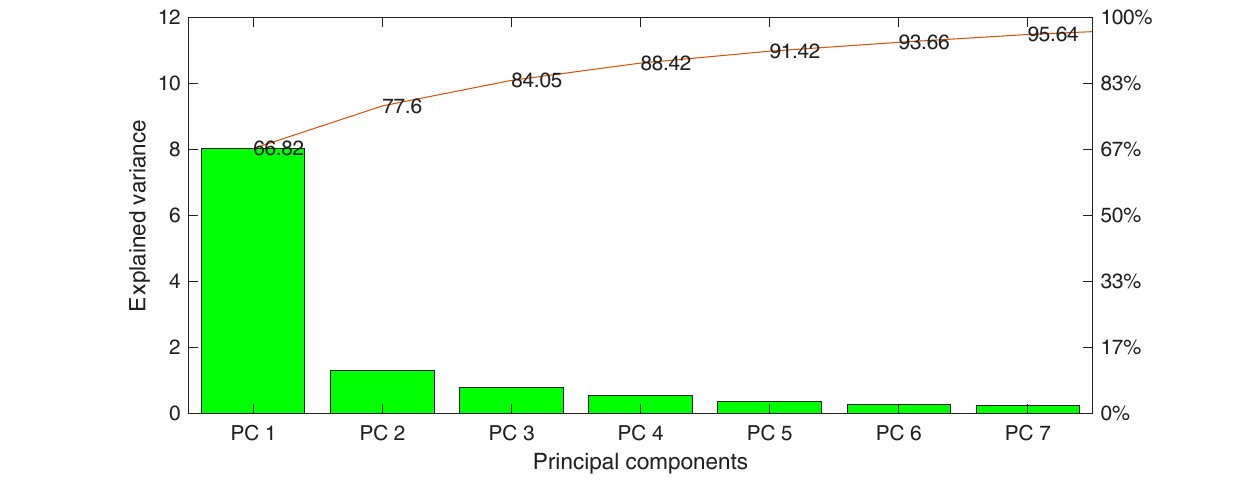}
\caption{Variance explained by the Principal Components.}  \label{VarExpl}
\end{figure}

\begin{figure}[!htb]
\centering
\includegraphics[trim=0cm 0cm 0cm 0cm, clip, width=1\textwidth]{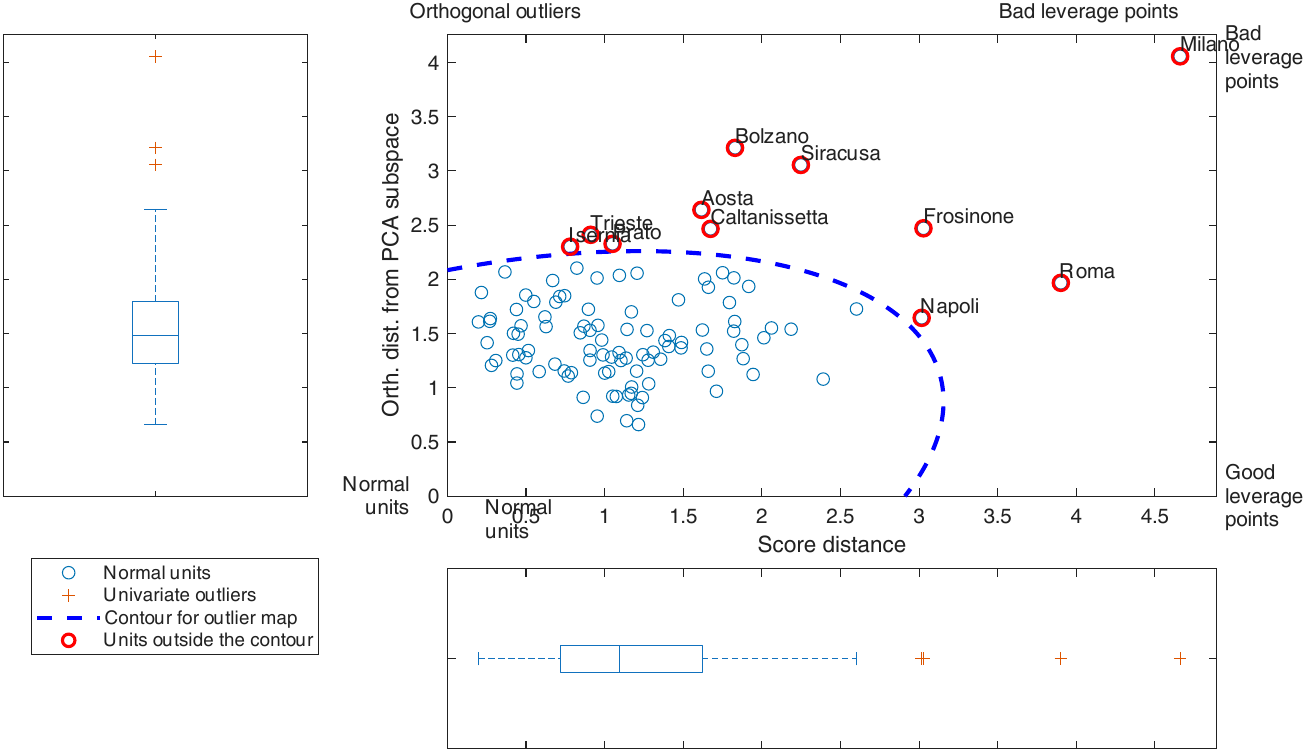}
\caption{Outlier map. The dashed contour is the inferential threshold for the combined score and orthogonal distances, computed using a Bonferronized 99\% confidence level. Units outside the contour are marked by red circles, while marginal univariate outliers in score or orthogonal distance are shown by crosses in the boxplots.}  \label{OutlierMap}
\end{figure}

The presence of three broad groups of variables suggested by the scatter plot matrix provides a useful reference for interpreting the subsequent results of the PCA, which we now apply to summarize these relationships and reduce dimensionality.
The call to function \mcode{pcaFS}
\begin{lstlisting}[firstnumber=8]
out = pcaFS(citiesItaly2024);
\end{lstlisting}
automatically enables us to obtain the plot of the total variation of the data explained by the principal components (Figure~\ref{VarExpl}),
the correlation of the 12 variables with the first two fitted principal components (Figure~\ref{NonSmartEigen}), the outlier map  (Figure~\ref{OutlierMap})  and the interactive biplot (Figure~\ref{APPbiplot}).
%
% In absence of optional arguments, in these figures colors are the FSDA defaults, being option \mcode{'colorBlindSafe'} set to \texttt{false}.
For backward compatibility, the default FSDA graphics retain the legacy colour
scheme when \mcode{'colorBlindSafe'} is set to \texttt{false}. In the revised
figures used for interpretation in this paper, the option
\mcode{'colorBlindSafe'} is set to \texttt{true} whenever grouped, signed, or
ordered quantities are displayed.
A detailed numerical output is also produced in the Command Window (Appendix B).
The output of the call to \mcode{pcaFS} is a \mcode{struct} (in this case named \mcode{out}) which contains as fields (in table and array format) all the information given in the Command Window and the principal component scores.

 \begin{figure}[!htb]
\centering
\includegraphics[trim=0cm 0cm 0cm 0cm, clip, width=1\textwidth]{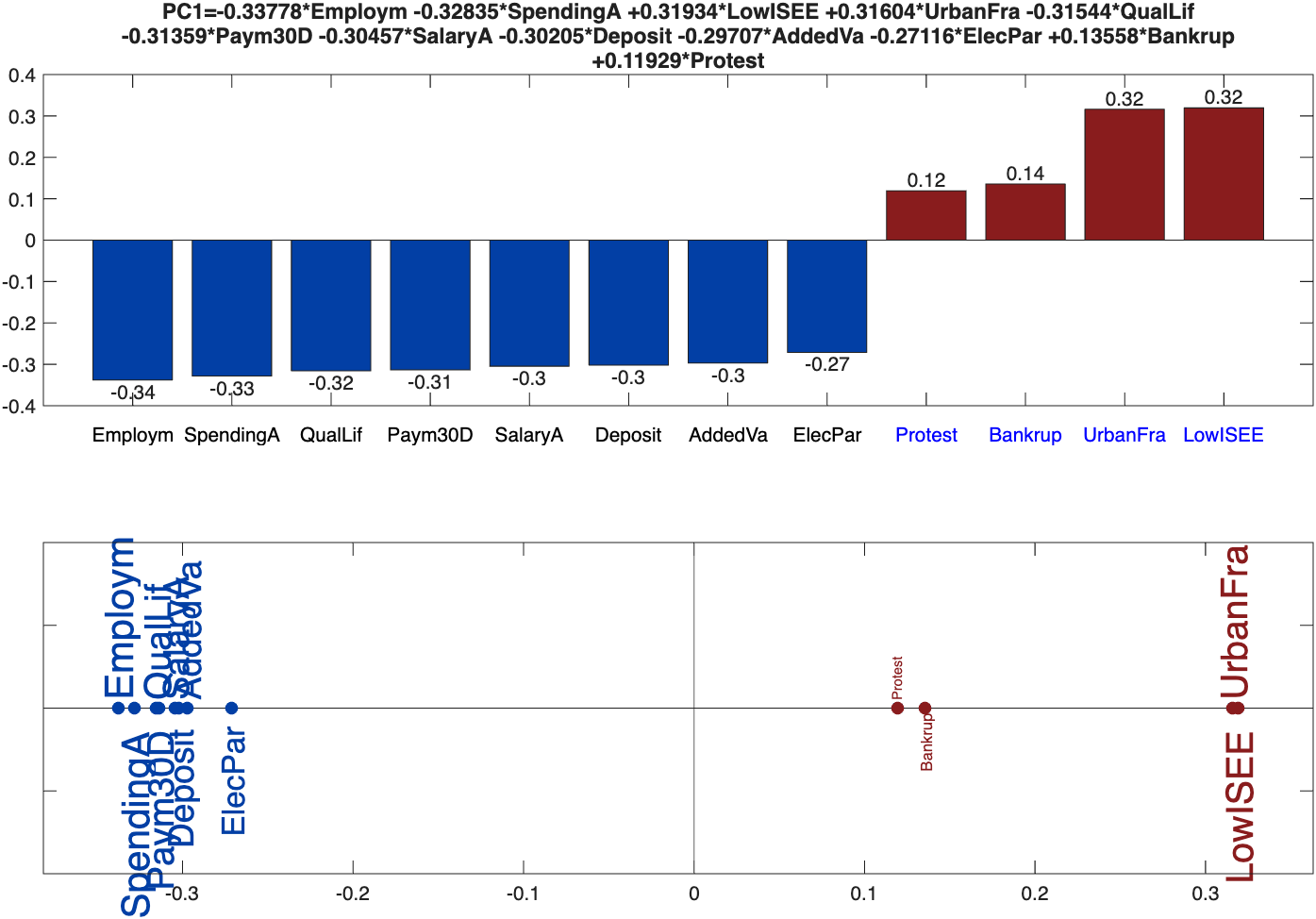}
\caption{``Smart'' representation of the ordered elements of the first eigenvector. The font size in the lower panel is proportional to the magnitude of the element. \mcode{'colorBlindSafe'} is set to \texttt{true}.}  \label{SmartEigen1}
\end{figure}

We now consider how much of the total variation in the data is explained by the principal components.  Figure~\ref{VarExpl}  shows that the first principal component explains almost 67\% of the total. With two principal components this rises to near 77\% and to 84\% with three. We can expect that the first two principal components will give a good representation of the majority of the data.

\begin{figure}[!htb]
\centering
\includegraphics[trim=0cm 0cm 0cm 0cm, clip, width=1\textwidth]{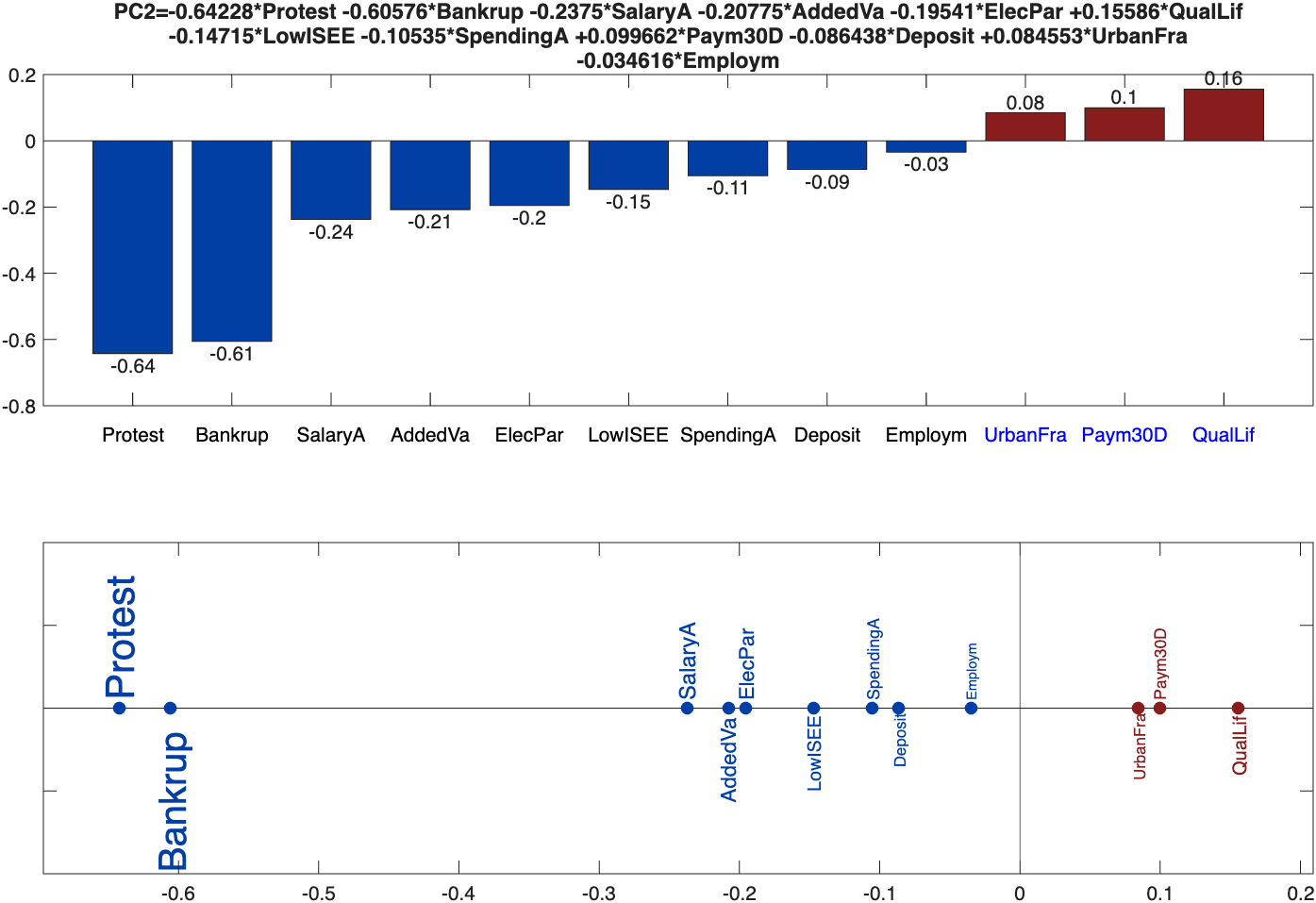}
\caption{Smart representation of the elements of the second eigenvector. \mcode{'colorBlindSafe'} is set to \texttt{true}.}  \label{SmartEigen2}
\end{figure}

If \mcode{pcaFS} is called with option pair \mcode{<'smartEVchart',true>} the plots with a ``smart'' representation of the elements of the first two eigenvectors are shown (Figures~\ref{SmartEigen1} and \ref{SmartEigen2}). In these figures, also option \mcode{'colorBlindSafe'} is set to \texttt{true} and the bars, stems, and variable labels are colored using the two extreme colors of the same HCL-based diverging palette used elsewhere in the paper: a blue hue for variables with a negative loading and a red hue for variables with a positive loading, replacing the fully saturated, non-colorblind-safe red/blue pair used by default by the function. The sign of each loading therefore remains immediately readable from color alone, while the ordering and magnitude of the loadings are conveyed, as before, by the horizontal position of the bars and by the font size of the variable names in the lower stem plot.

The upper panel of Figure~\ref{SmartEigen1}  gives,  in the title, the equation of  PC1 (using the order of the elements of the first eigenvector) with the ordered elements  of the first eigenvector in the lower panel.
The three groups already identified in Figure~\ref{spmAOE} are clearly recovered in this representation, in which the variables are ordered by the magnitude of their correlations with PC1, with names in a font size proportional to the absolute values. The three groups remain clear, with ``Protest'' and ``Bankruptcy'' being less extreme in location and with a low coefficient.
Figure~\ref{SmartEigen2}  shows the plots of the elements of PC2 with the corresponding equation.
Unlike PC1, which reflects the dominant socio-economic gradient, PC2 highlights contrasts within this structure, particularly separating variables such as ``Protest'' and ``Bankruptcy''. In fact,
%The variables ``Protest'' and ``Bankruptcy'' are important for defining this PC;
both  are remote from the centroid of the projections of the other variables and have appreciable weightings.

The interpretation of these plots is easy and provides insights into the structure of the data. Figure~\ref{NonSmartEigen}  shows the plots of the correlations between the variables and PC1 and PC2 in the order of the original data. It is possible, but not straightforward, to find some of the structure that is apparent in Figures~\ref{SmartEigen1} and  \ref{SmartEigen2}.

We now turn to the examination of the properties of the individual regions relative to the dominant structure, as revealed by the first two PCs.
The first plot we consider is an extended version of the ``outlier map'',  introduced by \citet{h+r+v:2005}, which is shown in Figure~\ref{OutlierMap}.

The plot summarizes aspects of the projection of the row points onto the two-dimensional plane formed by the PCs. The ``score distance'' is the Mahalanobis distance of the projected point from the centroid of the projections. The $y$ axis of the plot is the orthogonal distance of the point from the PCA subspace. Good leverage points have large score distances and small orthogonal distances. The inferential contour in the plane of score and orthogonal distances is based on the combined value of the two distances and uses a Bonferronized 99\% confidence level. Units outside the contour are therefore observations that are jointly extreme with respect to their projection on the PCA subspace and their residual distance from it.

More specifically, let \(z_i\) denote the standardized, row vector
associated with unit \(i\), let \(V_k\) be the matrix containing the first
\(k\) loading vectors and let
\(\Lambda_k=\mathrm{diag}(\lambda_1,\ldots,\lambda_k)\) be the diagonal
matrix of the corresponding eigenvalues. The score vector of unit \(i\) is
\(t_i=z_i V_k\). The score distance and the orthogonal distance used in the
outlier map are
\[
  \mathrm{SD}_i =
  \left(t_i \Lambda_k^{-1} t_i^\top \right)^{1/2},
  \qquad
  \mathrm{OD}_i =
  \left\|z_i-z_i V_k V_k^\top\right\|.
\]
The contour in Figure~\ref{OutlierMap} is obtained by applying a bivariate
rule to the vector of distances
\[
  d_i =
  \begin{pmatrix}
  \mathrm{SD}_i \\
  \mathrm{OD}_i
  \end{pmatrix}.
\]
If \(\widehat{\Sigma}_d\) denotes the covariance matrix of these two distances,
computed from the units used to estimate the PCA subspace, the combined
distance is
\[
  T_i^2 =
  d_i^\top \widehat{\Sigma}_d^{-1} d_i .
\]
The plotted contour is the level set
\[
  d^\top \widehat{\Sigma}_d^{-1} d
  =
  c_B,
  \qquad
  c_B =
  \chi^2_{2,\,1-\alpha_B},
  \qquad
  \alpha_B=\frac{1-0.99}{n},
\]
where \(n\) is the number of units screened. Thus the nominal 99\% level is
Bonferronized over the \(n\) units. Units satisfying \(T_i^2>c_B\) are
highlighted in red.

Figure~\ref{OutlierMap} shows that Milan is the most evident bad leverage point, with both a large score distance and a large orthogonal distance. Bolzano and Syracuse are mainly orthogonal outliers, whereas Rome, Frosinone and Naples have comparatively large score distances. The marginal boxplots remain useful for detecting units that are extreme in either distance separately, but the contour gives the bivariate outlier rule used to identify the units highlighted in red. The three groups indicated in the preceding three figures are not evident in this plot.

This section has illustrated how interactive graphical tools can be used to explore the structure of multivariate data through PCA. Starting from a simple scatter plot matrix, we progressively introduced ordering, correlation summaries, and dimension reduction to reveal dominant patterns and groupings among variables. The analysis shows that a small number of principal components capture most of the variability in the data and that their interpretation is facilitated by interactive visualizations that link scores, loadings, and correlation structures. These representations provide insight into both global trends and local deviations, and they motivate the use of biplots and related tools for a more detailed investigation of the relationships among individual observations and variables, which is the focus of the next section.

\section{Dynamic biplots and properties of individual regions}
\label{indivsec}

This section gives Step 3 of the workflow: using the dynamic biplot to connect variable interpretation, observation scores, confidence regions, and distance diagnostics. The controls shown in the screenshots are part of the implemented \mcode{biplotFS} interface and are also launched from \mcode{pcaFS}.

The biplot \citep{Gabriel1971Biplot} overlays a score plot with a  loading plot. Although the  biplot allows information on both samples and variables of a data matrix to be displayed in the same graphic, there is a problem in relating the scales of the two plots to obtain visual impact.
We build the biplot from the singular value decomposition (SVD) of the standardized deviation matrix $Z$ (or, equivalently, the mean–centred matrix $\tilde X$).  In what follows we work with $Z$ of size $n\times p$.  If instead one starts from $\tilde X$, replace references to the correlation matrix $R=Z^\top Z/(n-1)$ by the covariance matrix $S= \tilde X^\top \tilde X /(n-1)$.
The SVD of $Z$ is
\[
Z = U\,\Gamma\,V^\top .
\]
If the first two principal components capture most of the total variance, we approximate $Z$ by its best rank–2 approximation:
\[
Z \approx U_{(2)}\,\Gamma^*_{(2)}\,V_{(2)}^\top ,
\]
where $U_{(2)}$ is $n\times2$, $V_{(2)}=(v_1 \; v_2)$ is the $p\times2$ matrix containing the first two eigenvectors of $R$, and $\Gamma^*_{(2)}$ is the $2\times2$ diagonal matrix of the top two singular values of $Z$ -- the square roots of the leading eigenvalues of $Z^\top Z=(n-1)R$.  Writing instead
\[
\frac{Z}{\sqrt{n-1}}\approx U_{(2)}\frac{\Gamma^*_{(2)}}{\sqrt{n-1}}\,V_{(2)}^\top
\]
we obtain
\[
Z\approx \sqrt{n-1}\,U_{(2)}\frac{\Gamma^*_{(2)}}{\sqrt{n-1}}\,V_{(2)}^\top
=\sqrt{n-1}\,U_{(2)}\,\Gamma_{(2)}\,V_{(2)}^\top ,
\]
where $\Gamma_{(2)}=\Gamma^*_{(2)}/\sqrt{n-1}$.

A dynamic representation of row points and column points (arrows) of which the two previous representations are special cases is obtained by introducing two new parameters $\omega$ and $\alpha$ defined on $[0,1]$. The singular value decomposition is then written as
\[
Z \approx \left\{ (\sqrt{n-1})^\omega U_{(2)} \Gamma_{(2)}^\alpha \right\}
  \left\{ \Gamma_{(2)}^{1-\alpha}
    V_{(2)}^\top  (\sqrt{n-1})^{1-\omega} \right\}.
\]
In this case the $n$ row points can be represented by the $n \times 2$ matrix
\[
G=  (\sqrt{n-1})^\omega U_{(2)} \Gamma_{(2)}^\alpha,
\]
and the $p$ column points by the $2 \times p$ matrix
\[
H^\top  = \Gamma_{(2)}^{1-\alpha}  V_{(2)}^\top  (\sqrt{n-1})^{1-\omega}.
\]
A quite cryptic introduction to the biplot is given by \citet{v+rip:99}. Brief comments on scaling are on p.~336.
Two cases are of particular interest.\\
{\bf Case 1:} $\omega = 1$ and $\alpha=0$\\
In this case, row points represent standardized principal components $\sqrt{n-1}U=ZV\Gamma^{-1}$.
Arrows represent correlations between variables and principal components $\Gamma V$ (the loadings).
The length of the arrows is the communality of the variable (that is the proportion of the variance of a particular variable which is explained by the first two PCs).\\
{\bf Case 2:} $\omega = 1$ and $\alpha=1$\\
Now  row points represent principal components $\sqrt{n-1} U \Gamma=ZV$ and
arrows represent the eigenvectors $V$.

In both cases it is customary to add a unit circle.\\
{\bf Case 1.} The arrows close to the unit circle indicate the variables with communality close to 1.\\
\\
{\bf Case 2.} The arrows close to the unit circle indicate variables which play the most relevant part in the definition of the PC.

When
$\alpha=1$  interpoint distances in the biplot approximate Euclidean distances between observations in the original standardized space. Conversely, when
$\alpha=0$, interpoint distances approximate Mahalanobis distances, with respect to the correlation (or covariance) matrix, again up to truncation error. In the latter case the Euclidean distance $d_{ij}$ between two generic rows of matrix $\sqrt{n-1} U_{(2)}=(u_1^\top , \ldots, u_n^\top )^\top $ can be re-expressed. We use the notation of elementary vectors $e_i$,  $e_j$ (where $e_i$ is a vector of length $n$ with all elements equal to 0 except the $i$-th which is equal to 1), when
\begin{eqnarray*}
d_{ij}^2 & = &    ||u_i-u_j||^2= (n-1) (e_i-e_j)^\top  U_{(2)} U_{(2)}^\top  (e_i-e_j) \\
& = & (e_i-e_j)^\top    Z V_{(2)}   \Gamma_{(2)}^{-1} (Z V_{(2)}   \Gamma_{(2)}^{-1})^\top   (e_i-e_j) \\
& = &  (e_i-e_j)^\top    Z V_{(2)}   \Gamma_{(2)}^{-1} \Gamma_{(2)}^{-1}  V_{(2)}^\top   Z^\top   (e_i-e_j) \\
& \approx &  (e_i-e_j)^\top    Z R^{-1}   Z^\top  (e_i-e_j)
\end{eqnarray*}
because $V_{(2)}   \Gamma_{(2)} \Gamma_{(2)}  V_{(2)}^\top $ is the best rank 2 approximation of the correlation matrix $R$.
In short,
$\alpha$ controls whether emphasis is placed on interpreting relationships among variables ($\alpha=0$), or distances among observations $(\alpha=1)$.
The parameter
$\omega$, with  $\omega \in [0,1]$, determines how the factor $\sqrt{n-1}$ is distributed between rows and column points.
When $\omega=1$ (default choice) the scaling is entirely assigned to the row points. On the other hand when $\omega=0$
the scaling factor is assigned entirely to the column points, column vectors are emphasized, and row points are correspondingly shrunk.
Note that the parameter $\omega$ only redistributes the global factor $\sqrt{n-1}$
between row and column points and it does not change the geometric interpretation within rows or within columns -- only their relative size in the joint display.

\begin{figure}[!htb]
\centering
\includegraphics[width=\textwidth, trim=0 0 60 0, clip]{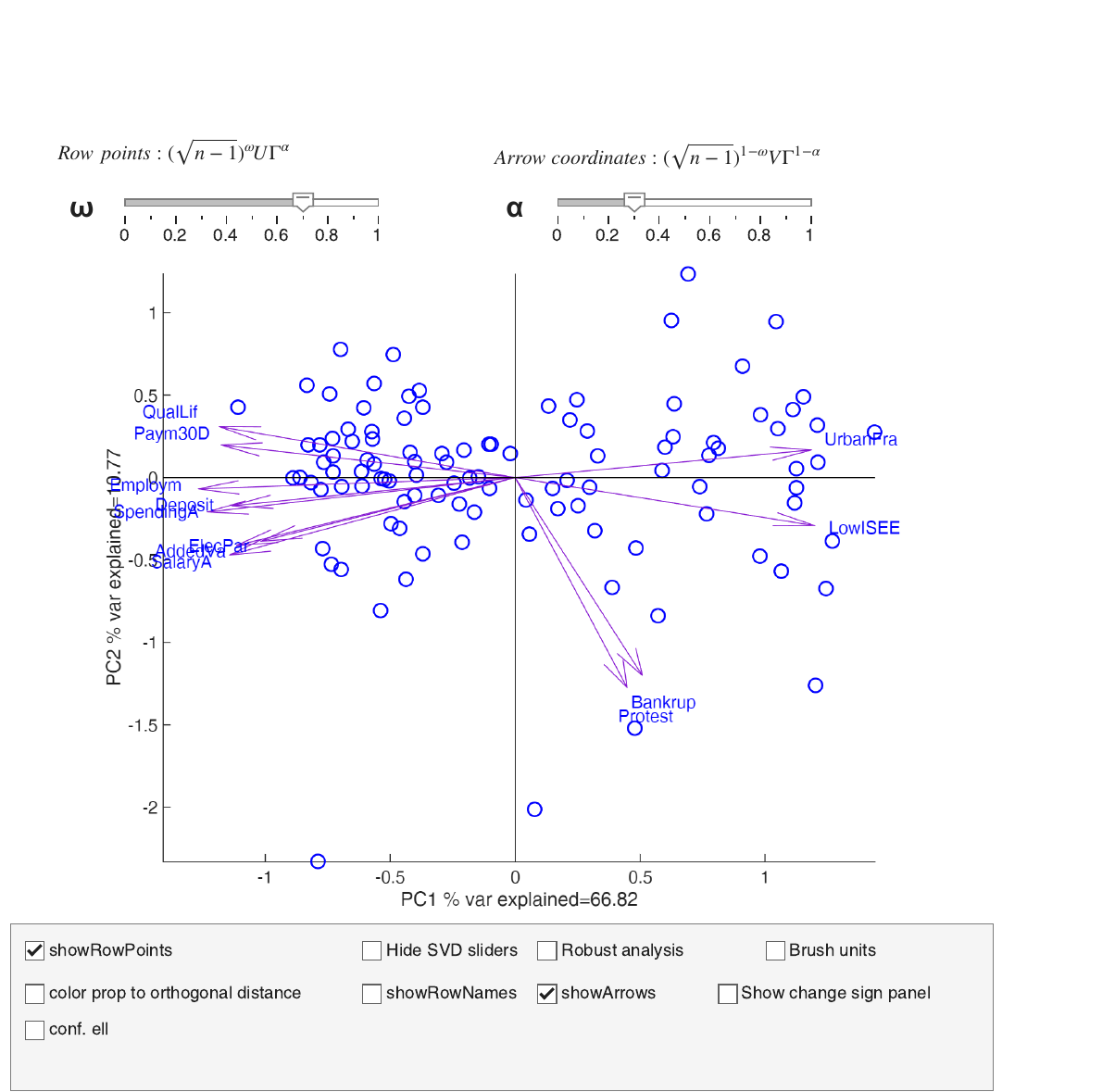}
\caption{Interactive biplot, showing row points and variables with labels using $\alpha=0.3$ and $\omega=0.7$. }  \label{APPbiplot}
\end{figure}

The geometric interpretation of the biplot depends critically on the choice of $\alpha$ and $\omega$. These parameters  shift emphasis between inter-point distances, relationships among variables and the relative scaling of row and column representations. We therefore provide an interactive GUI to explore these trade-offs in a dynamic and intuitive way. The two sliders at the top of Figure~\ref{APPbiplot} specify the values of $\alpha$ and $\omega$. The figure shows that there are also  options to show or hide score points and arrow labels, to control the percentage of displayed text labels, and to change the sign of the components.
When hovering on the control with the pointer an automatic datatip comes out which informs the user about the meaning of the associated control.

In the biplot of Figure~\ref{APPbiplot} we have taken $\alpha=0.3$ and $\omega=0.7$, values which we use in all the biplots in this paper as they lead, in this and many other examples, to a clear visual balance between the representations of variables and data points. The figure clearly shows the three groups of variables. Negative values of the first principal component are highly negatively correlated with variables indicating a good quality of life. The two variables positively correlated with PC1 indicate a poor quality of life. These variables are almost orthogonal to the two remaining variables, which have a small effect on the quality of life.

The outlier map plotted in Figure~\ref{OutlierMap} shows the scores of the units, that is their Mahalanobis distances from the centroid of the variables  in the plane of PC1 and PC2. It also shows the orthogonal distances of the units from the fitted plane. We extend the information from this plot, using the biplot to show both the location of the units in the space of PCs and their orthogonal distance from this plane.

 \begin{figure}[!htb]
\centering
\includegraphics[width=\textwidth, trim=0 0 60 0, clip]{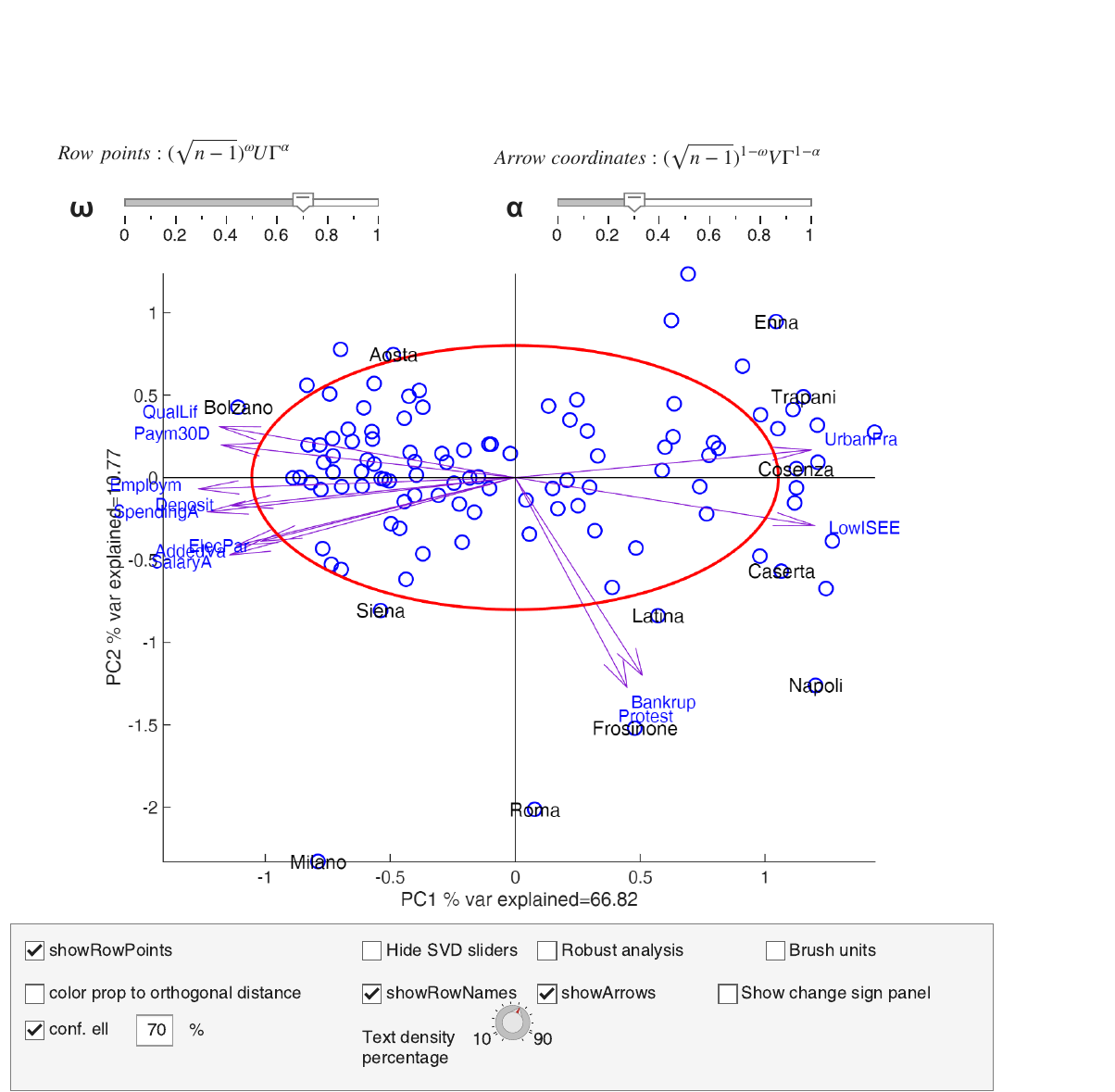}
\caption{Biplot with confidence ellipse.}  \label{APPconfell}
\end{figure}
In Figure~\ref{APPconfell}  we have added a confidence ellipse to the biplot of Figure~\ref{APPbiplot}. The ellipse is calculated under the assumption of normality with the semi-axes equal to the square root of the requested quantile of the $\chi^2$ multiplied by the square root of the variances of the scores depending on $\alpha$ and $\omega$.
The confidence level is specified in the edit field control located next to \mcode{conf. ell}. which automatically appears when the \mcode{conf. ell.} checkbox is selected.

Note that when $\alpha=0$ and $\omega=1$ the scores which are shown are standardized (their variance is equal to 1) therefore the ellipse becomes a circle.
 The names of the provinces are given only for those that lie outside the ellipse when check box \mcode{showRowNames} is selected. In this representation the three large cities Milan, Rome and Naples, with large negative values of  PC2 stand out, together with the smaller province of Frosinone, which lies physically between Rome and Naples.

\begin{figure}[!htb]
\centering
\includegraphics[trim=0cm 0cm 0cm 0cm, clip, width=0.48\textwidth]{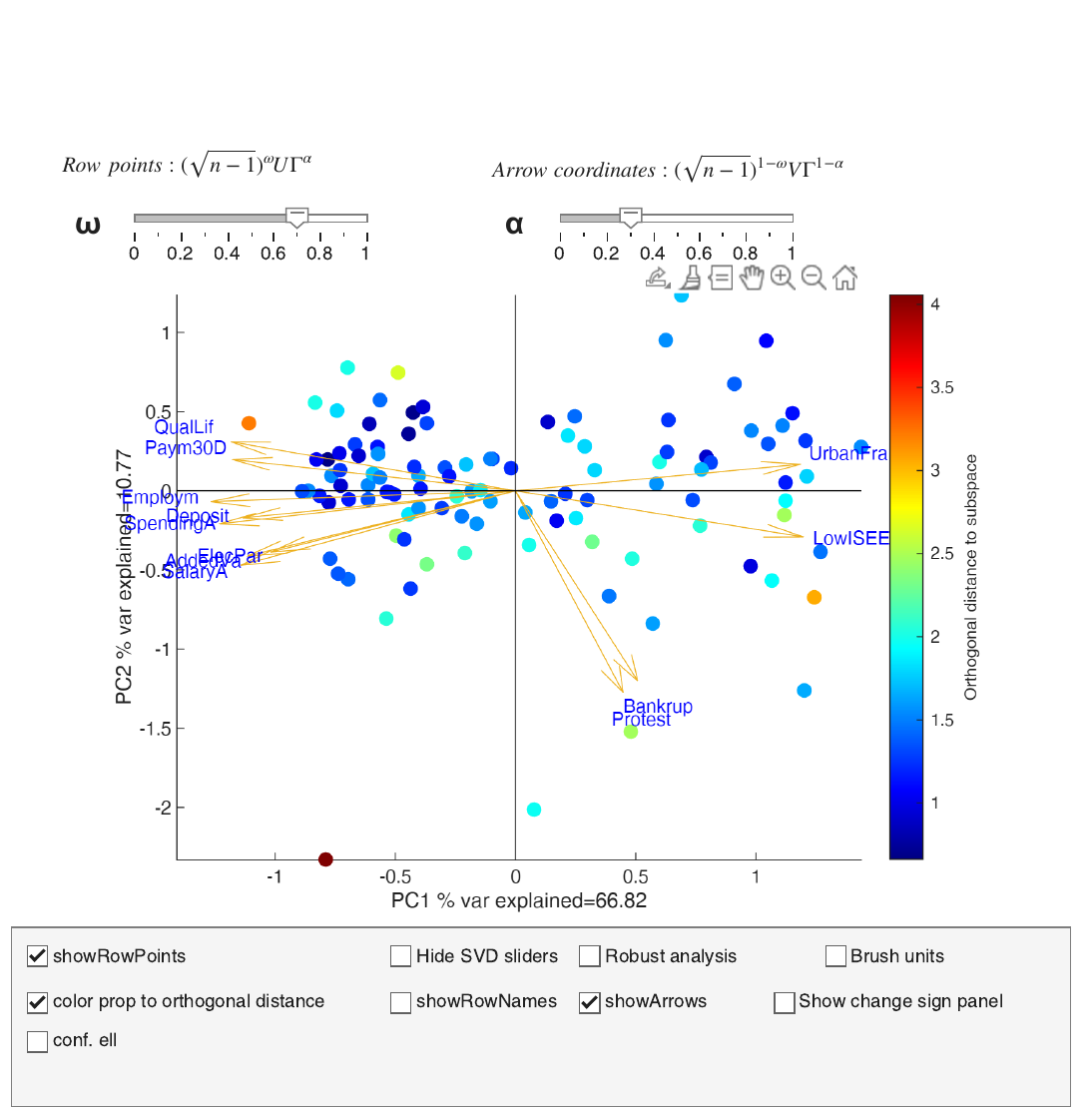}\hfill
\includegraphics[trim=0cm 0cm 0cm 0cm, clip, width=0.48\textwidth]{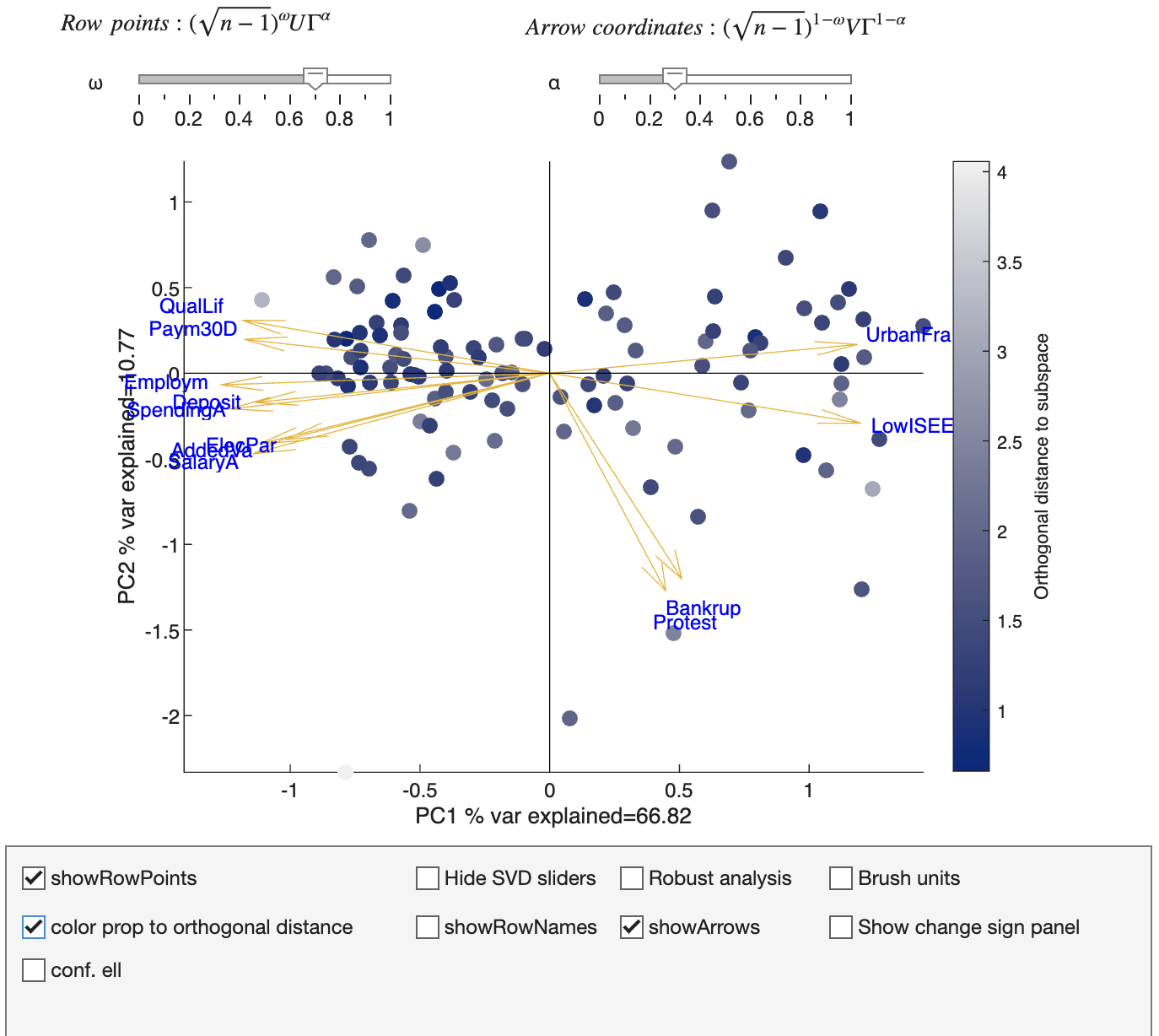}
\caption{Biplot with row points coloured according to the orthogonal distance
to the PCA subspace. The left panel shows the legacy palette
(\mcode{'colorBlindSafe'} set to \texttt{false}); the right panel shows the
revised colour-blind-safe palette (\mcode{'colorBlindSafe'} set to
\texttt{true}). The PCA geometry is unchanged in the two panels.}
\label{APPorth}
\end{figure}

To provide the information on the orthogonal distance from the plane the biplot of Figure~\ref{APPbiplot} must be augmented with the option \mcode{color prop to orthogonal distance}. 
Figure~\ref{APPorth} illustrates the effect of the new
\mcode{'colorBlindSafe'} option for a biplot in which row points are coloured
according to their orthogonal distance from the PCA subspace. The comparison
shows that the statistical information displayed in the plot is unchanged,
whereas the revised palette provides a more accessible representation of the
distance scale.

The predominance of colors at the lower end of the sequential scale indicates that the properties of nearly all provinces are well explained by the first two PCs. The three provinces at the high end of the scale are, in decreasing order of distance, Milan, Bolzano and Syracuse. Figure~\ref{APPorthRowNames}, shows the biplot with colour for the orthogonal distances and a percentage of names around 45\% as specified by the knob controller \mcode{Text density percentage}. Of course a percentage equal to 100 leads to a very  congested plot while a percentage close to zero  shows very  few labels.

The scatter plot matrix of Figure~\ref{spmAOE}, and the ensuing plots and the biplots,  all indicate that there are three groups of variables with high correlations within groups.  We now consider the effect of the replacement of the twelve variables with just three, one representative of each group. The variables we keep are Employment (Employm), representing all those positively correlated with QoL (see Table~\ref{table:QoL}), Urban Fragility (UrbanFra), representing negatively correlated variables and Protests per capita (Protest) to represent those with no correlation with quality of life.

 \begin{figure}[!htb]
\centering
\includegraphics[trim=0cm 0cm 0cm 0cm, clip, width=1\textwidth]{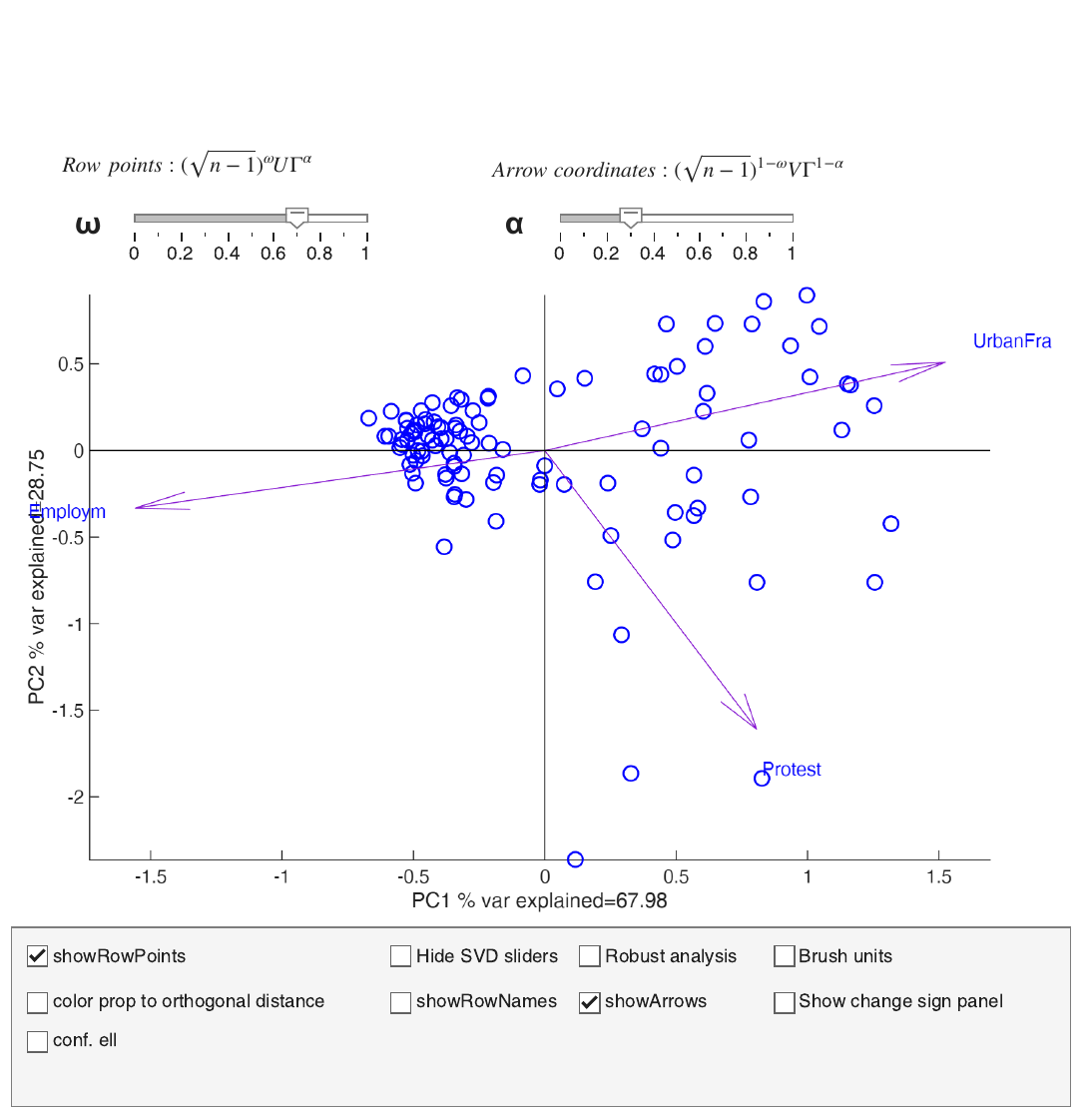}
\caption{Biplot based on just 3 variables.}  \label{APP3biplot}
\end{figure}

The resulting biplot obtained with the code\footnote{Function \mcode{biplotFS} enables us to obtain just the biplot. Of course, if all the other plots were needed, we could call \mcode{out=pcaFS(Xsel)}.}
\begin{lstlisting}[firstnumber=14]
Xsel=citiesItaly2024(:,{'Employm' 'Protest' 'UrbanFra'});
out=pcaFS(Xsel);
\end{lstlisting}
is shown in Figure~\ref{APP3biplot}. The reduction from 12 variables to 3 removes the visual clutter of too many arrows. It also sharpens the visualization of units on the biplot. Compared with Figure~\ref{APPorth}, Figure~\ref{APP3biplot}  indicates that the data contain a tight cluster of about 60 units with small negative values of PC1, with the others virtually all having more dispersed positive values of PC1.

%\section{Robustness}
\section{Robustness in PCA}
\label{robsec}

This section gives Step 4 of the workflow and demonstrates the robust-PCA branch of the implementation. The purpose is to show not only which observations are flagged by a robust fit, but also how the low-dimensional PCA structure changes as the trimming level is varied.

 \begin{figure}[!htb]
\centering
\includegraphics[trim=0cm 0cm 0cm 0cm, clip, width=1\textwidth]{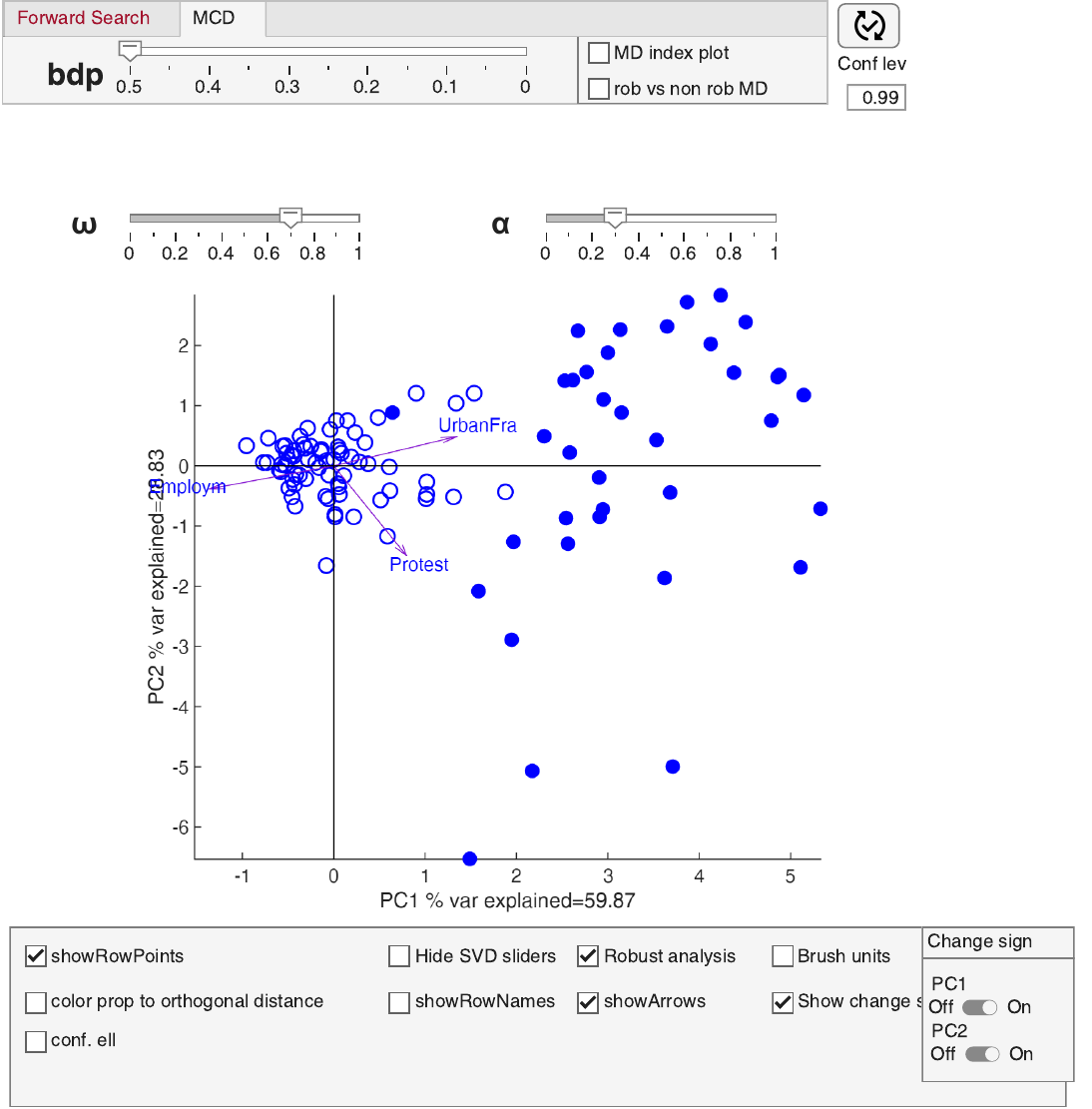}
\caption{MCD. Biplot based on just 3 variables using the units not declared as outliers with bdp = 0.5 and a simultaneous confidence level of 99\%. Note that in this case the sign of the first two PCs has been changed (see Change sign panel) in order to have the same layout as the other plots.}  \label{APP3mcd}
\end{figure}

\begin{figure}[!htb]
\centering
\includegraphics[trim=0cm 0cm 0cm 0cm, clip, width=1\textwidth]{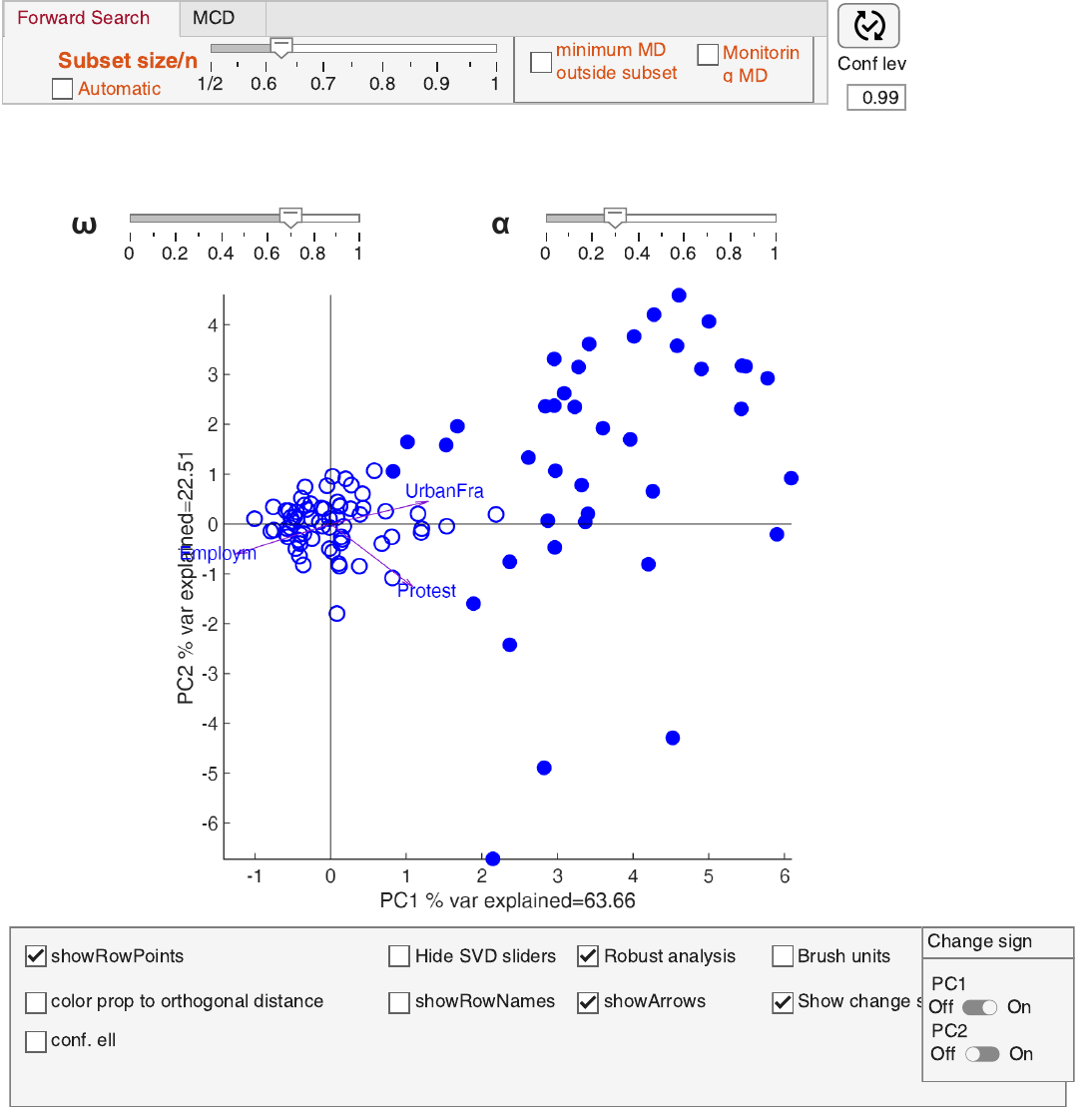}
\caption{Forward Search. Biplot based on just 3 variables. The highlighted units (filled circles) are those not belonging to the subset when $m=0.625n$.}  \label{APP3FS}
\end{figure}

We have already mentioned that the variables in PCA are scaled, often to have zero mean and unit variance. Any outliers in the original data will affect the scaling, perhaps appreciably. This suggests that robust methods for PCA should use hard trimming. That is, that some observations are given weight zero, while others are given weight 1. This is distinct from soft-trimming, or downweighting, such as S-estimation \citep{yohai:2019}, in which outlying observations are given weights appreciably less than one. One such hard trimming method is the ROBPCA algorithm of \citet{h+r+v:2005} which uses the minimum covariance determinant (MCD) estimator with a specified breakdown point (bdp) to find a central ellipsoid of outlier free data, to which PCA is applied.
%
% DP: to check if it is not too self-congratulatory

In our biplot GUI once the check box ``Robust analysis'' is selected a panel with two tabs named ``Forward Search'' and ``MCD'' automatically appears on the top of the GUI.
Figure~\ref{APP3mcd} shows the biplot for the very robust choice of 0.5 for the bdp of the MCD algorithm. The filled circles in the plot are outliers from the MCD fit at the 99\% simultaneous level.  This plot clearly shows the compact group of units seen in Figure~\ref{APP3biplot} and establishes that  about 35\% of the units  do not belong to this group. In the MCD literature it is common to show the index plot of robust Mahalanobis distances and to compare  robust and non-robust distances. These two plots can be obtained simply by selecting the associated check boxes \mcode{MD index plot} and  \mcode{rob vs non rob MD} (see top right part of Figure~\ref{APP3mcd}).

While ROBPCA provides a well-established and computationally efficient approach to robust PCA, its output corresponds to a single choice of trimming level and yields a fixed robust fit. In contrast, the FSDA framework allows robustness to be explored dynamically, by examining how the fitted structure evolves as the effective subset size changes. This perspective emphasizes robustness not as a single estimate, but as a continuum of solutions indexed by the degree of trimming.
Hard trimming using the Forward Search (FS) estimates the parameters of the model from  subsets of  increasing size $m$ from the data. A crucial point is that, for each value of $m$, the model is refitted, the subset of size $m+1$ being chosen as the units with the $m+1$  smallest Mahalanobis distances. In this way observations can enter or leave the subset as $m$ increases.
% DP - addition
This sequential inclusion mechanism provides a natural way to study the stability of the principal components and to assess the influence of individual observations. In particular, it allows the analyst to distinguish between structural features of the data and effects driven by a small number of influential units, something that is difficult to assess from a single robust fit alone.
 Figure~\ref{APP3FS} shows the biplot for the FS when  $m = 0.625n$, which is close to the  size of the outlier free subset identified by the MCD
 (if we select $m = 0.65n$ we  obtain exactly the same plot as Figure~\ref{APP3mcd}). This biplot has very similar principal component directions to that of Figure~\ref{APP3mcd}. This stability suggests that the dominant structure is not driven by a small subset of observations. At the same time, the FS provides additional information by showing how and when individual units enter the model, offering a richer diagnostic perspective than a single robust fit.
\citet{Robook:2024} describe numerous robust methods, with a focus on the use of the FS in regression. Chapter 5 of \citet{arc:2004} describes and exemplifies the FS for PCA, but without the advantages of the modern interactive graphics that are the subject of this paper. In the FS literature it is common to show forward plot minimum Mahalanobis distances and to monitor the MD as the subset size increases. This last plot is an extension of the plots which compare the robust and non robust Mahalanobis distances. These two plots can be obtained simply by selecting the associated check boxes \mcode{minimum MD outside subset} and \mcode{monitoring MD} (see top right part of Figure~\ref{APP3FS}).

The structure revealed by our robust analyses is of a compact group of units, with many outliers. If the breakdown point of the robust fit is reduced, outliers are included in the observations used to estimate the parameters and, consequently, the group may not be evident. Figure~\ref{APP3mcd025} shows the biplot when the bdp of ROBPCA is appreciably reduced from 0.5 to 0.25. The principal components do not change much, so that the scatter of the points is similar to that of Figures~\ref{APP3mcd} and ~\ref{APP3FS}. However, only 4 outliers are identified. An automatic method for the detection of groups in the data with such a low bdp would fail.

Note that the proportion of variance explained by the MCD estimator varies with the choice of the bdp and, in general, so will the principal components and the biplot. This follows because these quantities are computed using only those observations that are not identified as outliers. Similarly, in the FS approach, the explained variance depends on the specific subset of units included at each step of the procedure.

%DP
The combination of interactive graphics and forward monitoring therefore provides insight that goes beyond standard robust PCA implementations, supporting a more informed and transparent exploration of multivariate structure.
Within FSDA, robustness is thus treated as an exploratory dimension rather than a fixed modelling choice: hard trimming methods and interactive graphical tools allow users to examine how principal components and data structure evolve as the degree of robustness changes.

\section{Linked brushing and geography}
\label{brushsec}

This section gives Step 5 of the workflow. Brushing is used to define subsets in the PCA display, and the  selected indices are propagated to scatter plot matrices and maps.
In the interactive session the analyst can draw these selections with the mouse.
The replication scripts launch the same linked displays used in the paper and document
the commands required to reproduce the computational workflow. The final appearance
of the brushed figures depends on the manual selections described in
Section~\ref{sec:replication}.

\begin{figure}[!htb]
\centering
\includegraphics[trim=0cm 0cm 0cm 0cm, clip, width=1\textwidth]{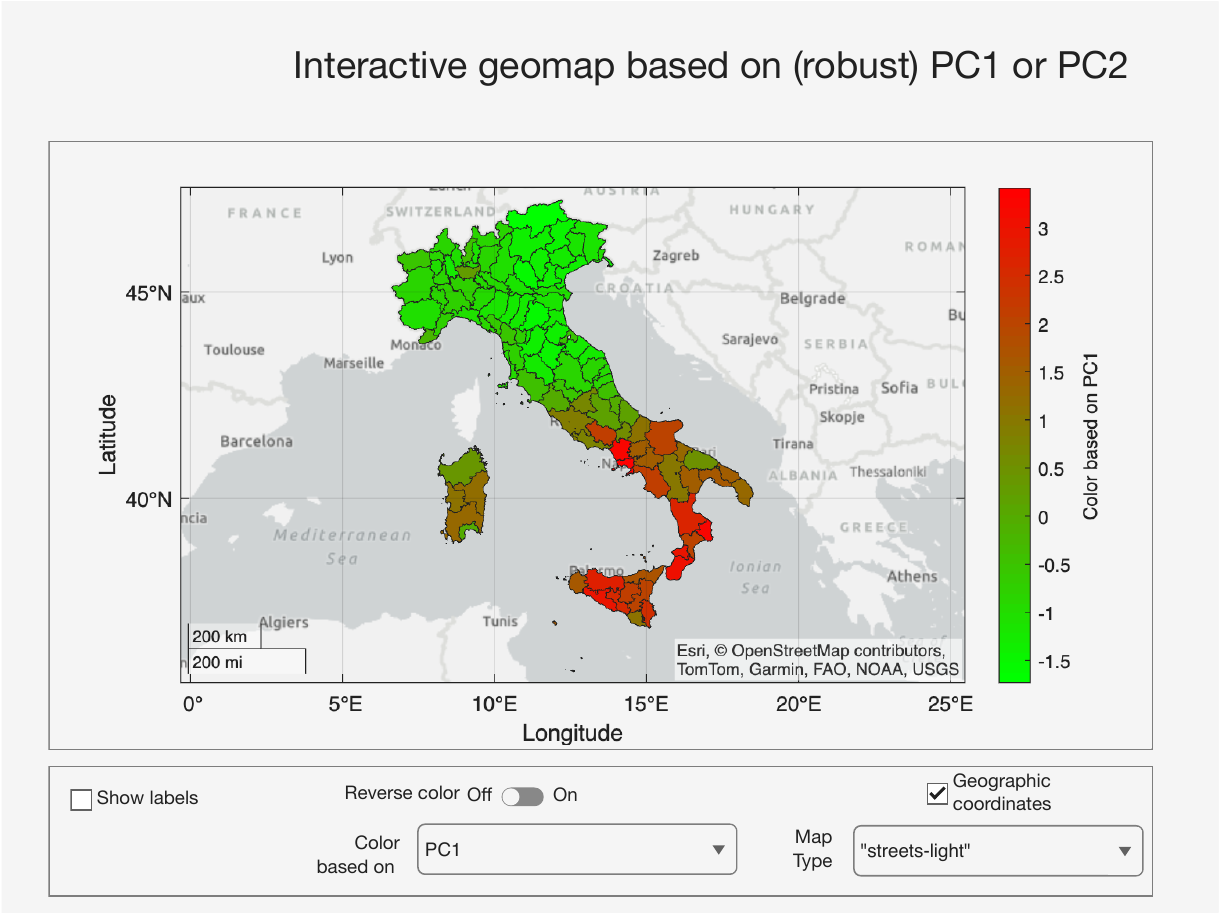}
\caption{Interactive geomap with colour based on the values of the first PC. Type of map is ``streets-light''. The map for the second PC is exhibited by clicking on the \mcode{Color based on} Drop Down control.}  \label{APPgeomap}
\end{figure}

We continue with the analysis including only the three variables and now tie our results to the geography of Italy.
Function \mcode{pcaFS} also allows the user to supply the vector of latitudes and longitudes of the statistical units of the sample or the corresponding geographical shapes.
When the instruction \mcode{load citiesItaly2024.mat} is invoked the table \mcode{citiesItaly2024} which is loaded into memory contains inside \mcode{Properties.Userdata}
\begin{lstlisting}[firstnumber=10]
  citiesItaly2024.Properties.UserData
\end{lstlisting}
a cell array with two elements.
\begin{lstlisting}[numbers=none]
ans =
  1 x 2 cell array
    {107 x 13 table}    {107 x 2 double}
\end{lstlisting}
The first is a geographic table (called a geotable in MATLAB) that is a table with the field named \mcode{Shape} containing the shapes of the provinces and the second is a matrix with two columns containing respectively the latitude and longitude coordinates.
The call
\begin{lstlisting}[numbers=none]
ShapeFile=citiesItaly2024.Properties.UserData{1};
out=pcaFS(Xsel,'ShapeFile',ShapeFile);
\end{lstlisting}
unlike  the  plots that we have seen so far, produces an interactive geomap (see Figure~\ref{APPgeomap}). As a default the colouring is based on the first PC but it can be changed to the second PC. Similarly it is possible in an interactive way to hide or show labels, specify the type of map, the coordinates (satellite or not) the map type and reverse the colour.

In order to have additional information of the geographical position of the units which occupy particular areas in the biplot we use brushing and linked plots.

 \begin{figure}[!htb]
\centering
\includegraphics[trim=0cm 20cm 0cm 0cm, clip, width=1\textwidth]{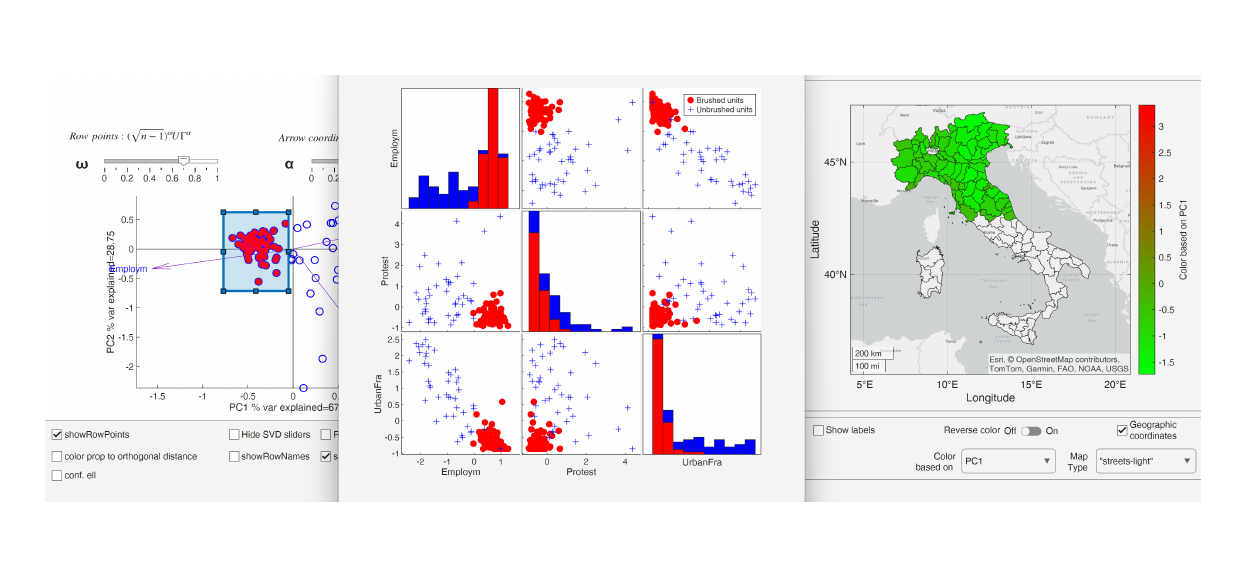}
\caption{Brushing the most compact group which corresponds (with the exception of Milan) to all the provinces located in the north-central part of Italy.}  \label{brush1}
\end{figure}

Our initial example of brushing is illustrated in Figure~\ref{brush1}, which we treat in some detail. We start from the non-robust biplot of Figure~\ref{APP3biplot}, that is with three variables and all provinces. We use the mouse to select the group of provinces that form the compact group. The centre panel of the figure shows scatter plots of the brushed units as filled highlighted circles. They indeed have a much  smaller scatter in all panels of the scatter plot matrix than the non-brushed units. Notice that the legends  are clickable, in the sense that clicking on the item in the legend, it is possible to hide or show the corresponding units in the scatter plot matrix. The other plot, which automatically appears after brushing, is the one given in the right-hand panel, which shows the position of the brushed provinces on the map of Italy.  With the exception of Milan,  all the provinces  are located in the north-central part of Italy.

\begin{figure}[!htb]
\centering
\includegraphics[trim=0cm 20cm 1cm 0cm, clip, width=1\textwidth]{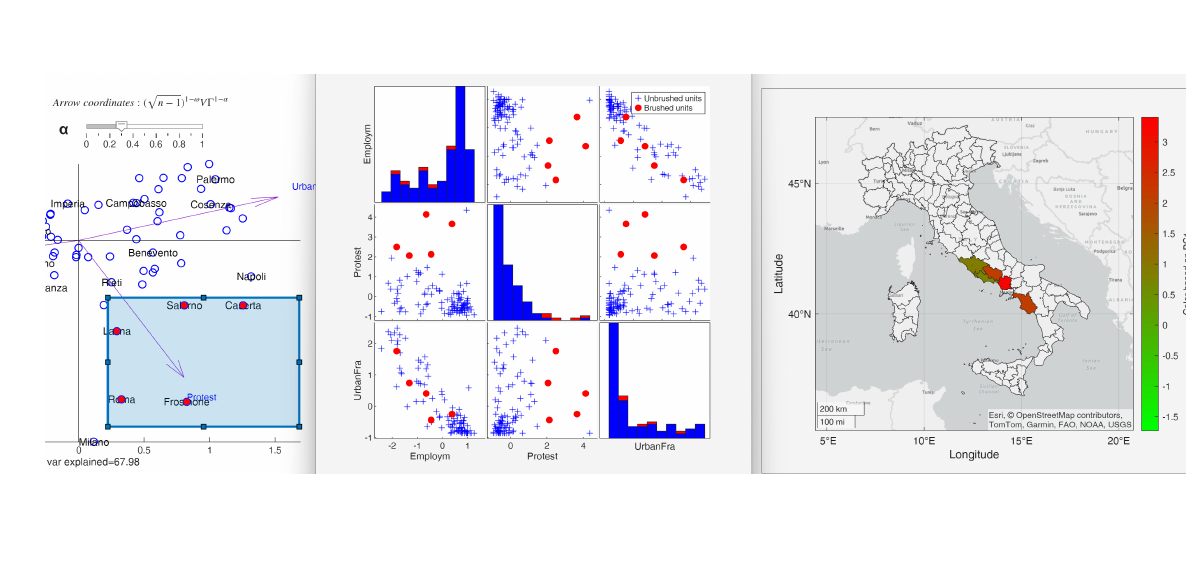}
\caption{Brushing the units with the largest values of protest (in the lower right part of the biplot).}  \label{brush2}
\end{figure}

The geographical outcome of this analysis could not be clearer. We continue by brushing the units with the  largest value of protest. Figure~\ref{brush2} shows that these lie in the lower right-hand quadrant of the biplot. The scatter plot matrix in the centre panel of Figure~\ref{brush2} shows that high values of protest are not related to employment or urban fragility. The map of these provinces shows that they all lie along the coast on either side of Naples. Note, however, that Naples is very close to the provinces selected for brushing in the left part of Figure \ref{brush2}.

\begin{figure}[!htb]
\centering
\includegraphics[trim=0cm 22cm 1cm 0cm, clip, width=1\textwidth]{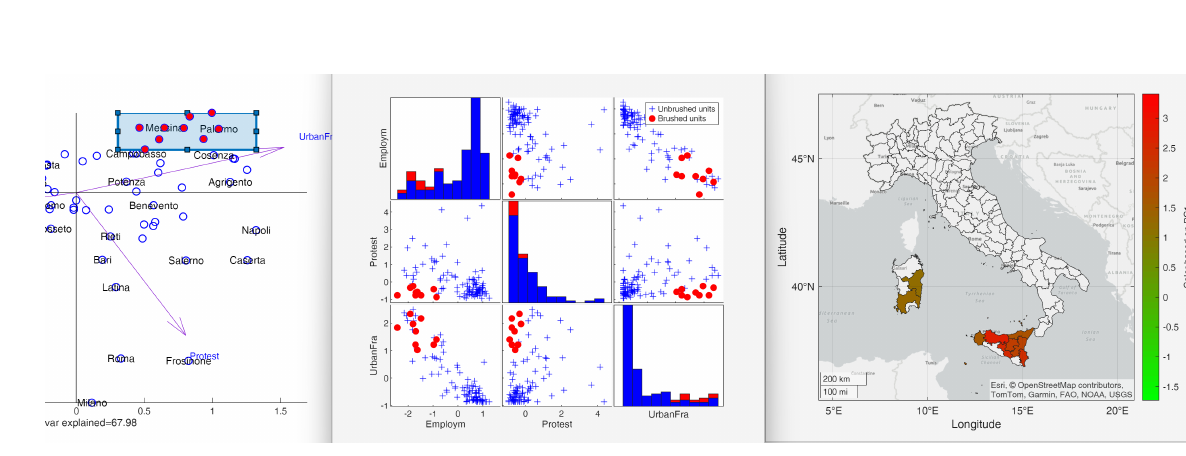}
\caption{Brushing the units with the largest values of urban fragility (in the upper right part of the biplot). }  \label{brush3}
\end{figure}

As a final analysis  of the data using three variables, we brush on the units with the largest values of urban fragility, which lie in the top-right hand quadrant of the biplot. The right-hand panel of Figure~\ref{brush3} shows that all these provinces lie in the south of Italy. The provinces  of Sicily tend to have  colors closer to the high end of the map scale than those of the provinces of Sardinia, in general indicating a lower quality of life.

In conclusion, we used three similar brushes on the biplot coming from the analysis including all 12 variables. This led to the identification of all except two of the provinces  shown in the PCA with just three variables. This provides a confirmation that the analysis based on the three variables is  representative of the analysis based on all  12 variables.

\begin{figure}[!htb]
\centering
\includegraphics[trim=0cm 20cm 1cm 0cm, clip, width=1\textwidth]{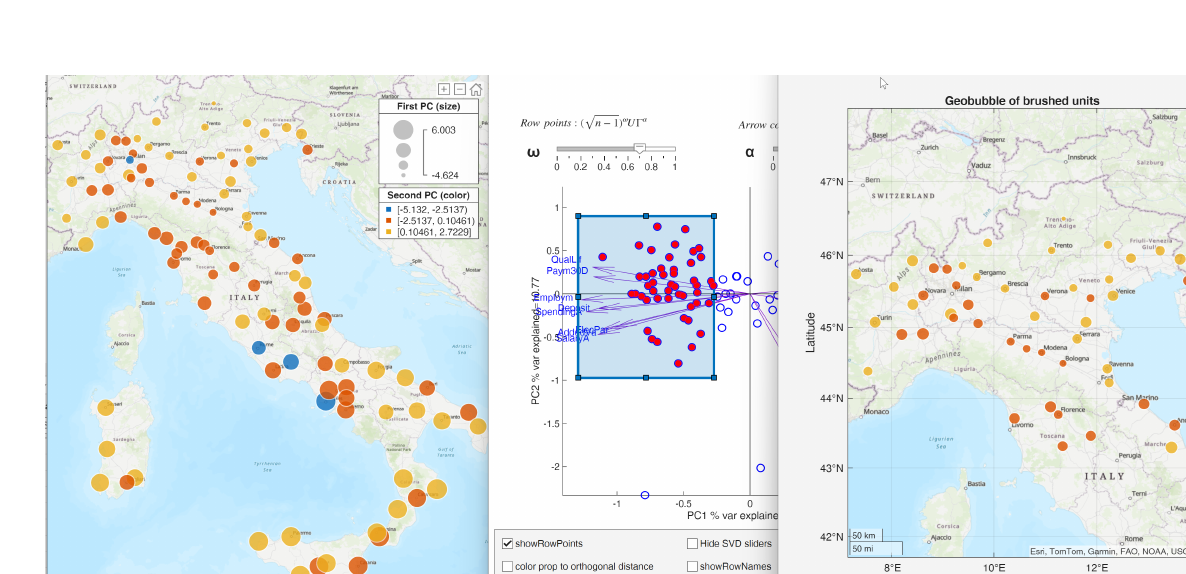}
\caption{Brushing in the space of first two PCs using all the 12 variables reveals the location of the units with the most negative values of the first PC.}  \label{APPLatLongBrush}
\end{figure}

\newpage

As a further analysis we now show what can be obtained  if  the latitude and longitude are included in the call to \mcode{pcaFS}.
The code below
\begin{lstlisting}[numbers=none]
LatLong=citiesItaly2024.Properties.UserData{2};
Latitude=LatLong(:,1);
Longitude=LatLong(:,2);
out=pcaFS(citiesItaly2024,'Latitude',Latitude,'Longitude',Longitude);
\end{lstlisting}
automatically adds, to the previous plots,  a geobubble plot, where the size of the bubble depends on the value of the first PC and the colour on the value of the second PC (see left panel of Figure~\ref{APPLatLongBrush}). When latitude and longitude are supplied, once a brushing is performed, the geobubble  showing the selected units (in addition to the scatter plot matrix) is automatically shown on the screen.
For example,  Figure~\ref{APPLatLongBrush} shows that once the provinces with the most negative values of the first PC are selected (with the exception of Milan at the bottom of the biplot, see Figure~\ref{APPconfell}), the corresponding geobubble of the selected units reveals that these units all belong to the northern central part of Italy. Specifically, the coastal towns of Liguria, including Genoa, are not in the group with the smallest values of PC1.

\section{Three-dimensional projection diagnostics}
\label{3sec}

\begin{figure}[!htb]
\centering
 % trim=left bottom right top
\includegraphics[trim=0cm 17cm 0cm 1.7cm, clip, width=1\textwidth]{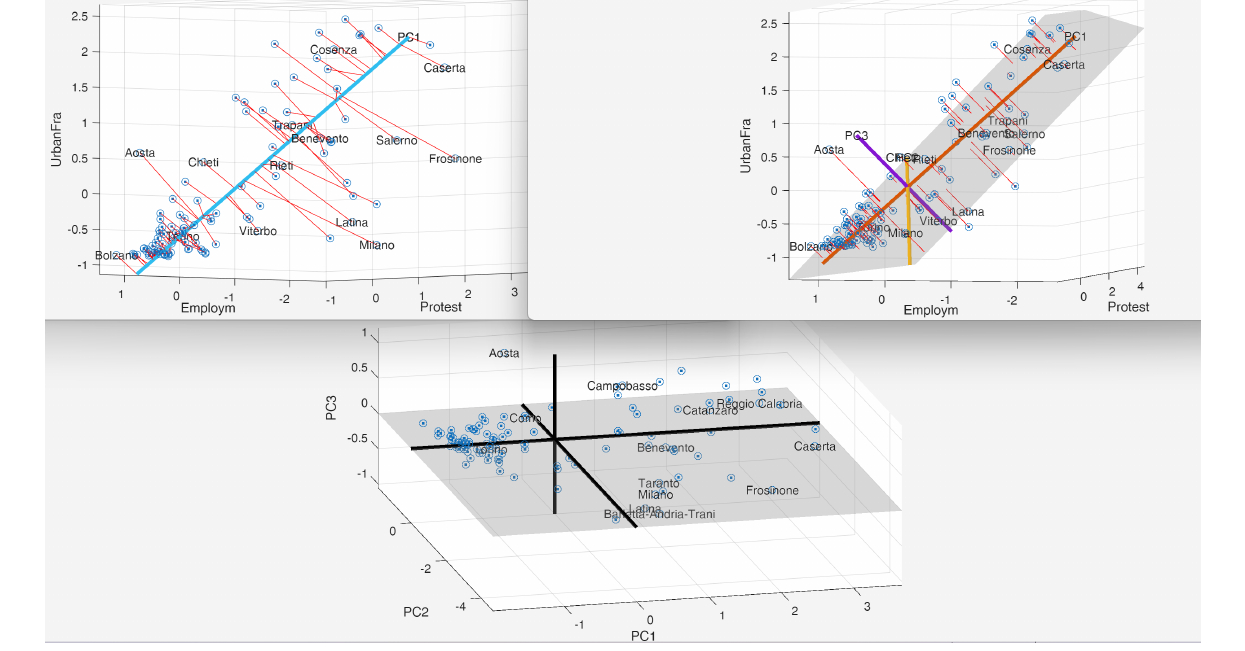}
\caption{3D geometry. Top left: projection in the plane of PC1, top right projection in the plane of the first two PCs.
Lower panel, plot in the space of the 3 PCs. Note that Aosta is not well represented in the space of the first two PCs.}  \label{Proj}
\end{figure}

\begin{figure}[!htb]
\centering
 % trim=left bottom right top
\includegraphics[trim=0cm 22cm 0cm 1.7cm, clip, width=1\textwidth]{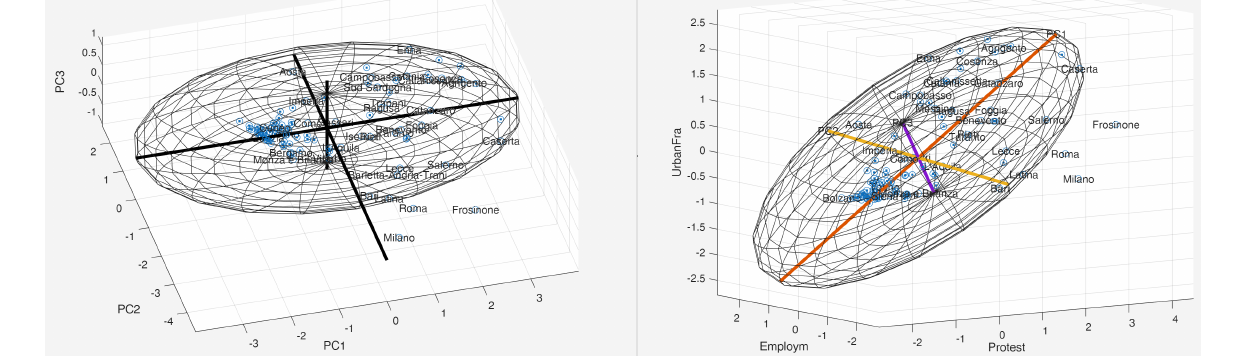}
\caption{3D geometry with 99\% ellipsoids. Left panel: ellipsoid in the space of the 3 PCs. Right panel: original 3D ellipsoid;
Milan, Rome and Frosinone are outside the confidence ellipsoid. Note that these three towns showed up in Figure~\ref{APP3mcd025}.}  \label{ProjEll}
\end{figure}

In this section we describe our routine \mcode{pcaProjection} which enables plotting of both the points in the original space and their respective projections in the low dimensional space together with a confidence ellipsoid in the two spaces.
The code below
\begin{lstlisting}[numbers=none]
pcaProjection(Xsel)
\end{lstlisting}
produces Figures~\ref{Proj} and \ref{ProjEll}. All these plots allow rotation. Therefore, a static printed view does not capture all the different angles of view.
Figure~\ref{Proj} illustrates different projections of the data: the top-left panel shows the projection onto PC1 alone, the top-right panel displays the projection onto the plane spanned by the first two PCs, and the third plot represents the data in the three-dimensional space of the first three PCs. This comparison highlights that Aosta is not well represented in the space of the first two principal components, since its structure becomes apparent only when the third component is taken into account.

Figure~\ref{ProjEll} compares the ellipsoid in the original and transformed space.
The left panel displays the confidence ellipsoid in the three–dimensional space spanned by the first three principal components, providing a geometric summary of the joint variability of the observations. The right panel shows the original 99\% confidence ellipsoid in the same space, superimposed with the observed data points. It is evident that Milan, Rome  and Frosinone lie outside the ellipsoid, indicating atypical behaviour relative to the main data structure captured by the first three PCs.

%\section{Replication materials and availability}
%\label{sec:replication}
%
%The replication material can be found in github at this address: \url{https://github.com/UniprJRC/PCAinteractive}.
%The root \texttt{README.md} describes this folder structure, the required \proglang{MATLAB} version and toolboxes, the \pkg{FSDA} version used for the paper, and the license information for the software and data.

\section{Replication material}
\label{sec:replication}

The replication material for this paper is distributed with the submission and
is also available from the public GitHub repository
\url{https://github.com/UniprJRC/PCAinteractive}. In response to the JSS
replication requirements, the material has been reorganized around a single
commented \proglang{MATLAB} script, \texttt{replication.m}, located in the
root directory of the repository. This file is the primary entry point for
reproducing the analyses in the paper.

The script \texttt{replication.m} is divided into labelled sections following
the order of the manuscript. It first checks that the required \pkg{FSDA}
functionality is available and loads the data set \texttt{citiesItaly2024.mat}.
It then reproduces the scatter plot matrix, the classical PCA output, the
dynamic biplot examples, the robust PCA analyses, the linked brushing and
geographical displays, the three-dimensional projection diagnostics, and the
appendix figures. Thus the computational workflow is documented and launched from a single script, in a single
file rather than being split across several figure-specific scripts.

The interactive operations described in the paper, such as brushing, linking,
dynamic updating of biplots, map interaction, and manual rotation of
three-dimensional figures, require either a \proglang{MATLAB} Desktop session
or \proglang{MATLAB Online}. The replication script launches the same
interactive displays used in the paper and contains comments indicating the
manual actions needed to reproduce the corresponding screenshots. Thus the
computational workflow is reproducible from a single script, while the final
visual appearance of some interactive figures depends on the user interaction
described in the relevant sections of the manuscript.

The data used in the paper are the 2024 quality-of-life indicators for the 107
Italian provinces. In the \proglang{MATLAB} script the data are loaded from
\texttt{citiesItaly2024.mat}, which is distributed with the required version of
\pkg{FSDA}. To facilitate use outside \proglang{MATLAB}, the repository also
contains the file \texttt{data/citiesItaly2024.geojson}. This file includes
the 12 variables used in the PCA analysis, the geographical contours of the
107 provinces, and the latitude and longitude information used by the linked
map displays. The auxiliary file \texttt{data/readInteractivePCA.R} reads the GeoJSON file
in \proglang{R} and creates three data frames: \texttt{Xeco}, containing the
12 quality-of-life variables; \texttt{XLatLong}, containing latitude and
longitude; and \texttt{Xgeo}, containing the geographical variables.

The graphical commands in \texttt{replication.m} use the colour-blind-safe
option of the new \pkg{FSDA} graphics whenever grouped, signed, or ordered
quantities are displayed. For example, the scatter plot matrix is generated by
\texttt{spmplot} with option \mcode{'colorBlindSafe'} set to \texttt{true},
and the PCA examples call \texttt{pcaFS} with the same option. Commented
alternative commands using the legacy palette are retained only to document
the difference between the previous and revised graphical defaults.

For convenience, the repository also contains an \texttt{extra/} directory.
This directory collects the earlier figure-specific scripts and the
corresponding \texttt{.ipynb} notebooks, which are retained only as optional
inspection and teaching material. They are not required for reproducing the
paper: the reference replication file is the single commented script
\texttt{replication.m} in the root directory.

\section{Summary, limitations, and extensions}
\label{futuresec}

This paper documents new \pkg{FSDA} functionality for interactive and robust PCA graphics in \proglang{MATLAB}. The contribution is software integration. The implemented workflow combines classical PCA output, dynamic biplot scaling, score and orthogonal-distance diagnostics, robust hard trimming through MCD and the Forward Search, linked brushing, geographical displays, and three-dimensional projection diagnostics. The Italian provinces case study illustrates how these components are used together: pairwise plots identify variable groups, classical PCA summarizes the dominant socio-economic gradient, robust analyses reveal the stability of this structure under trimming, and linked brushing connects PCA subsets to geography.

There are also limitations. The principal methodology remains linear PCA, so nonlinear structures may require methods outside the scope of the current implementation. The interactive layer relies on \proglang{MATLAB} graphics handles, UI components, and callbacks, which limits full compatibility with GNU Octave; the computational core is more portable than the interactive interface. Finally, color and interaction choices must be documented carefully because screenshots alone cannot fully convey an interactive analysis. The replication scripts therefore document the commands used to launch the same
interactive displays and provide reproducible code for the non-interactive parts of the
analysis, while the final appearance of some brushed or rotated figures depends on the
user interaction described in Section~\ref{sec:replication}.

Natural extensions include nonlinear dimension reduction methods such as Isomap \citep{tenenbaum2000global} and Locally Linear Embedding \citep{roweis2000nonlinear}, and autoencoder-based nonlinear PCA \citep{hinton2006reducing}. Such extensions would require new diagnostics because the biplot geometry used here is linear. Nevertheless, the same design principles -- common output structures, linked selections, robust diagnostics, and script-level reproducibility -- could be reused.

\bibliographystyle{plainnat}
%\bibliography{robustnew10Dec2014,PCAV4-extra}
\bibliography{robustnew10Dec2014}

\newpage

\appendix
\renewcommand{\thefigure}{A\arabic{figure}}
\renewcommand{\theHfigure}{A\arabic{figure}} % <-- important for hyperlinks
\setcounter{figure}{0}

%\section{Additional Figures}
%

%\renewcommand{\thefigure}{A\arabic{figure}}

\section{Further Figures}
\label{sec:further_figures}

In this appendix we present extra figures that complement, or contrast with, those in the main body of the paper.

\begin{figure}[!htb]
\centering
\includegraphics[width=1\textwidth,clip=true]{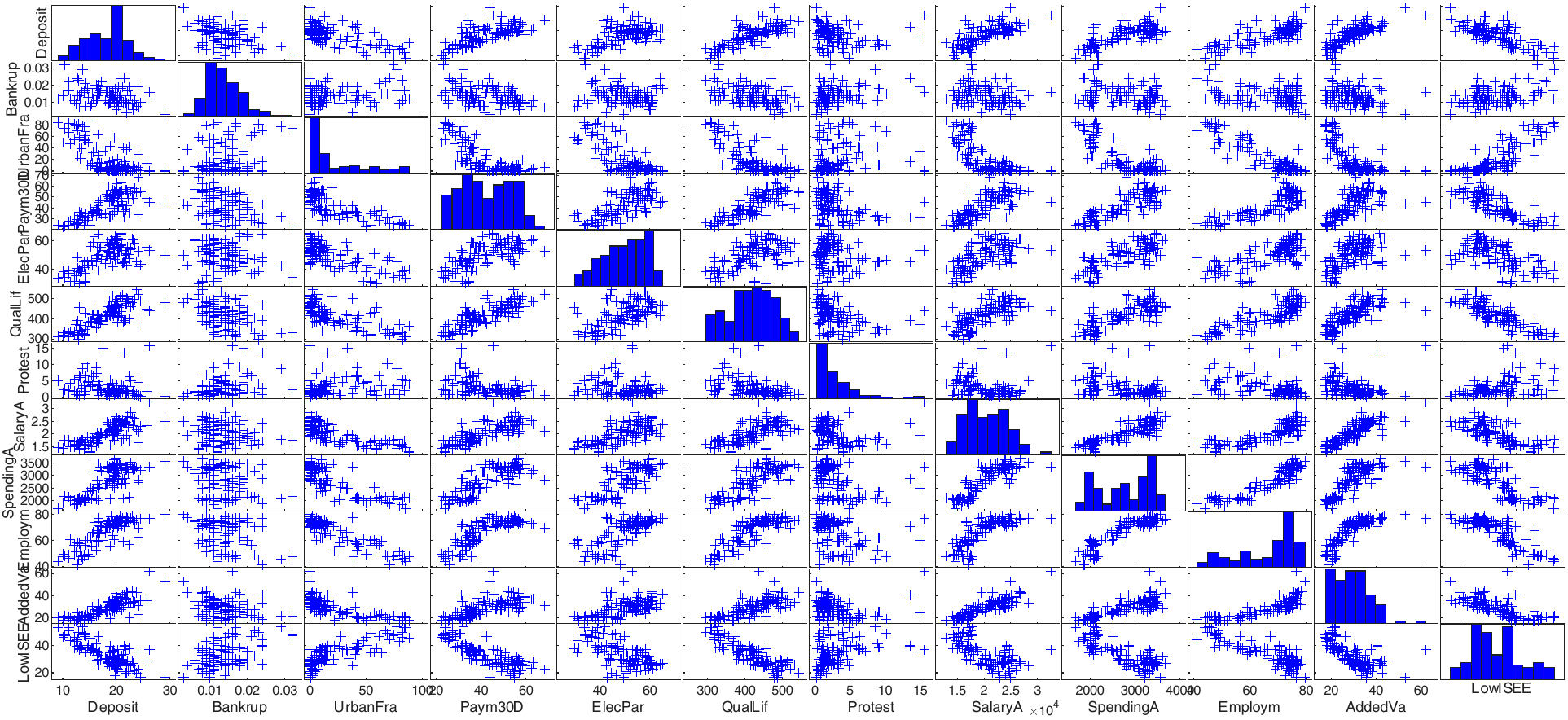}
\caption{Original standard scatter plot matrix with rows ordered as in the original data.}
\label{A1}
\end{figure}

Figure~\ref{A1} is the scatter plot matrix of the data in the order of presentation in \S\ref{datasec}. The grouping of the variables which was so evident in Figure~\ref{spmAOE} is not indicated by this plot. It is also not strongly indicated if the data in the original order are plotted using the modified scatter plot matrix. For these data, the ordering is crucial in suggesting the grouping structure.

 \begin{figure}[!htb]
\centering
\includegraphics[trim=0cm 0cm 0cm 0cm, clip, width=1\textwidth]{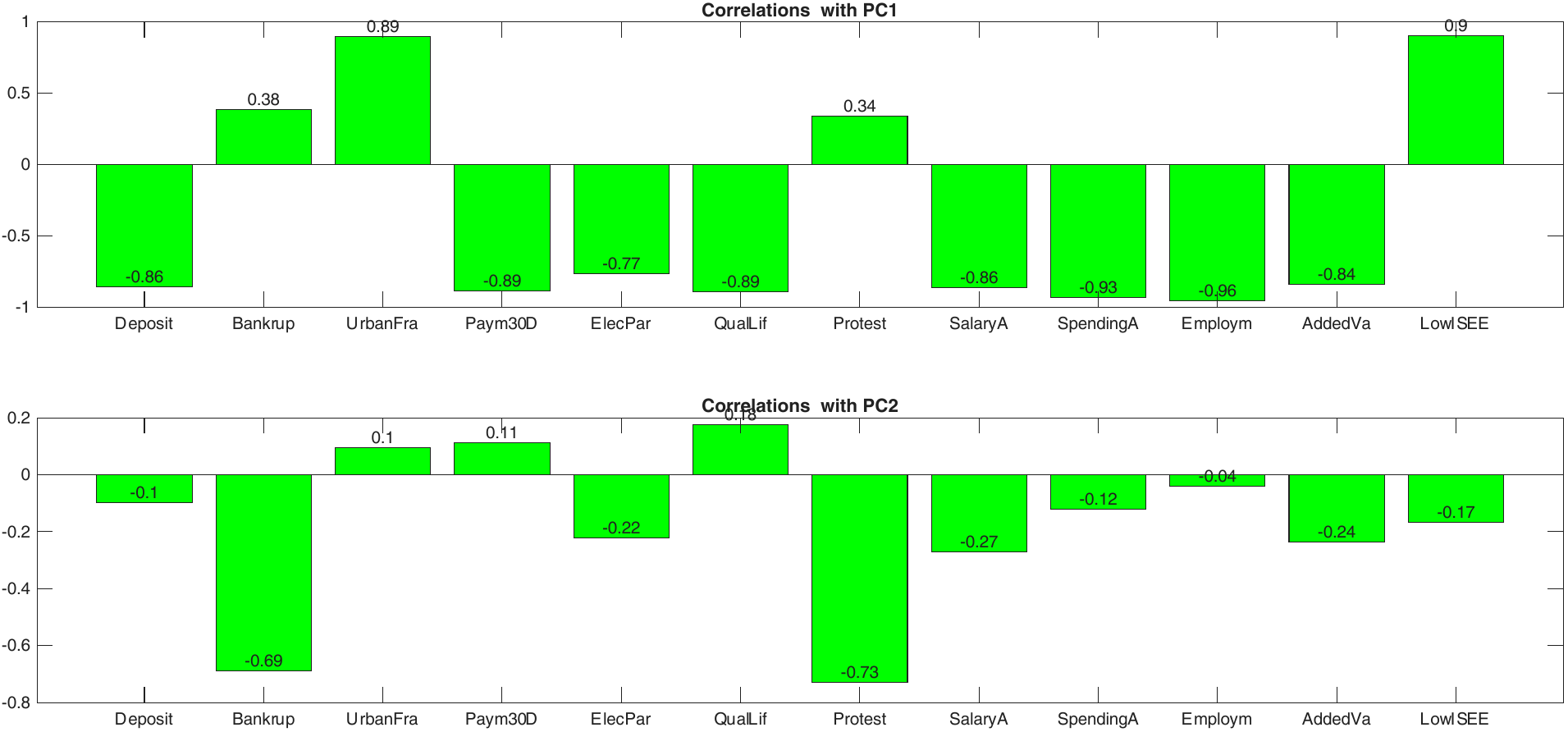}
\caption{Traditional representation of the correlations with PC1 and PC2. Option \mcode{'colorBlindSafe'} is not used and it is therefore  set by default to \texttt{false}.}  \label{NonSmartEigen}
\end{figure}

We now turn to the structure of the PCs. Figures~\ref{SmartEigen1} and \ref{SmartEigen2} show a ``smart'' representation of the elements of the first two eigenvectors. The elements are ordered by their contributions to the two PCs (so the order is different in the two plots). The conventional plot is shown in Figure~\ref{NonSmartEigen} in which the correlations of the variables with the two PCs is plotted in the order of the variables in Table~\ref{table:QoL}. As we remark in \S\ref{datasec}, it is possible, but not easy, to see some of the structure of the data in this plot. In particular, there are two variables with strong positive correlations with PC1 and two other variables with strong negative correlations with PC2. However, the lower panels of Figures~\ref{SmartEigen1} and \ref{SmartEigen2} contribute much more information.

 \begin{figure}[!htb]
\centering
\includegraphics[trim=0cm 0cm 0cm 0cm, clip, width=1\textwidth]{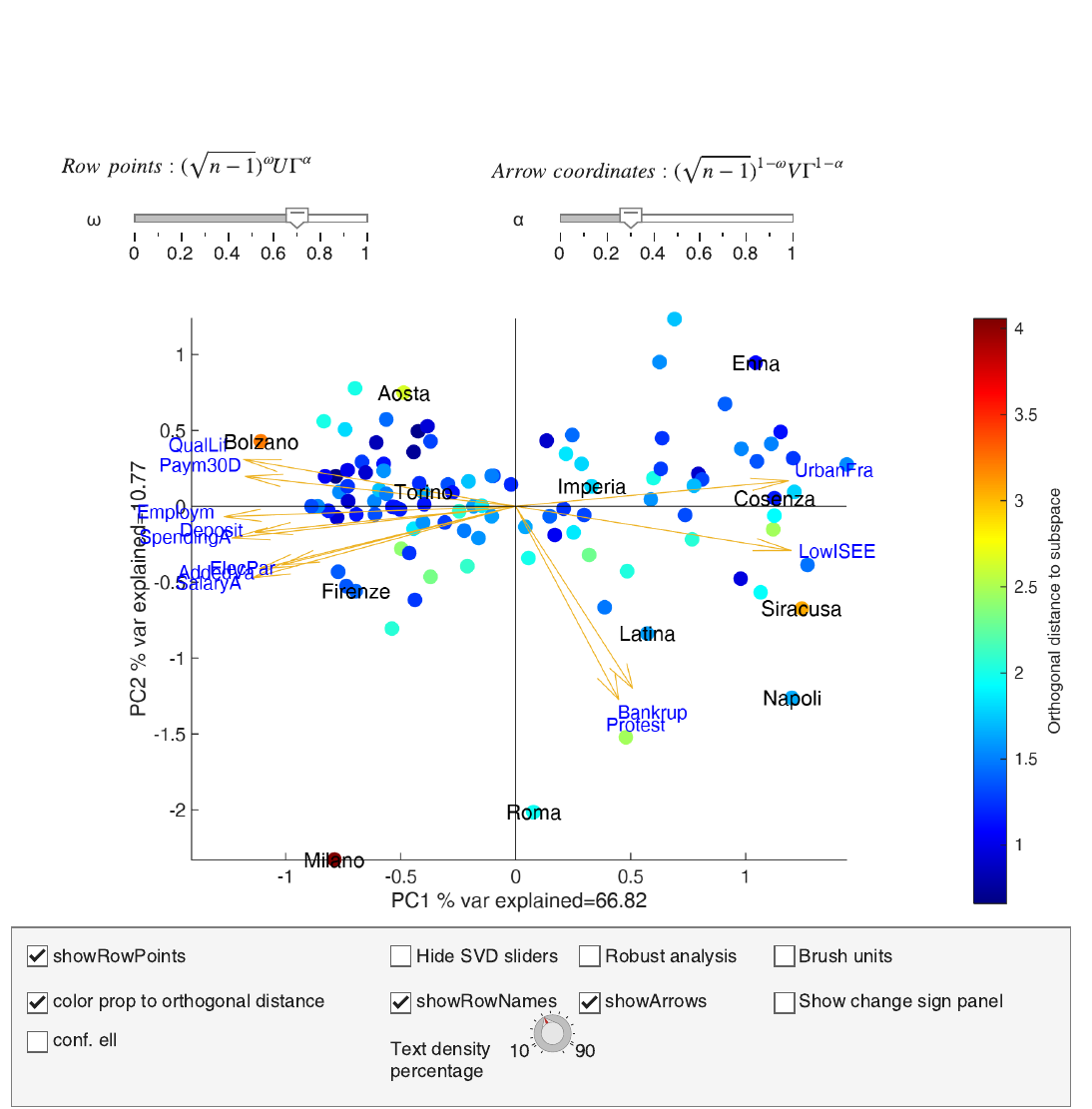}
\caption{Biplot with colour proportional to the orthogonal distance;  text density label around 45\%. Option \mcode{'colorBlindSafe'} is not used and it is therefore  set by default to \texttt{false}}  \label{APPorthRowNames}
\end{figure}

Figure~\ref{APPorth} introduced the use of colour in the biplot in order to express the distance of the units from the PC1-PC2 plane. Figure~\ref{APPorthRowNames} adds the names of the units to this plot. The percentage of names is around 45\% as specified by the knob controller \mcode{Text density percentage} which automatically appears once \mcode{showRowNames} checkbox is selected.

 \begin{figure}[!htb]
\centering
\includegraphics[trim=0cm 0cm 0cm 0cm, clip, width=1\textwidth]{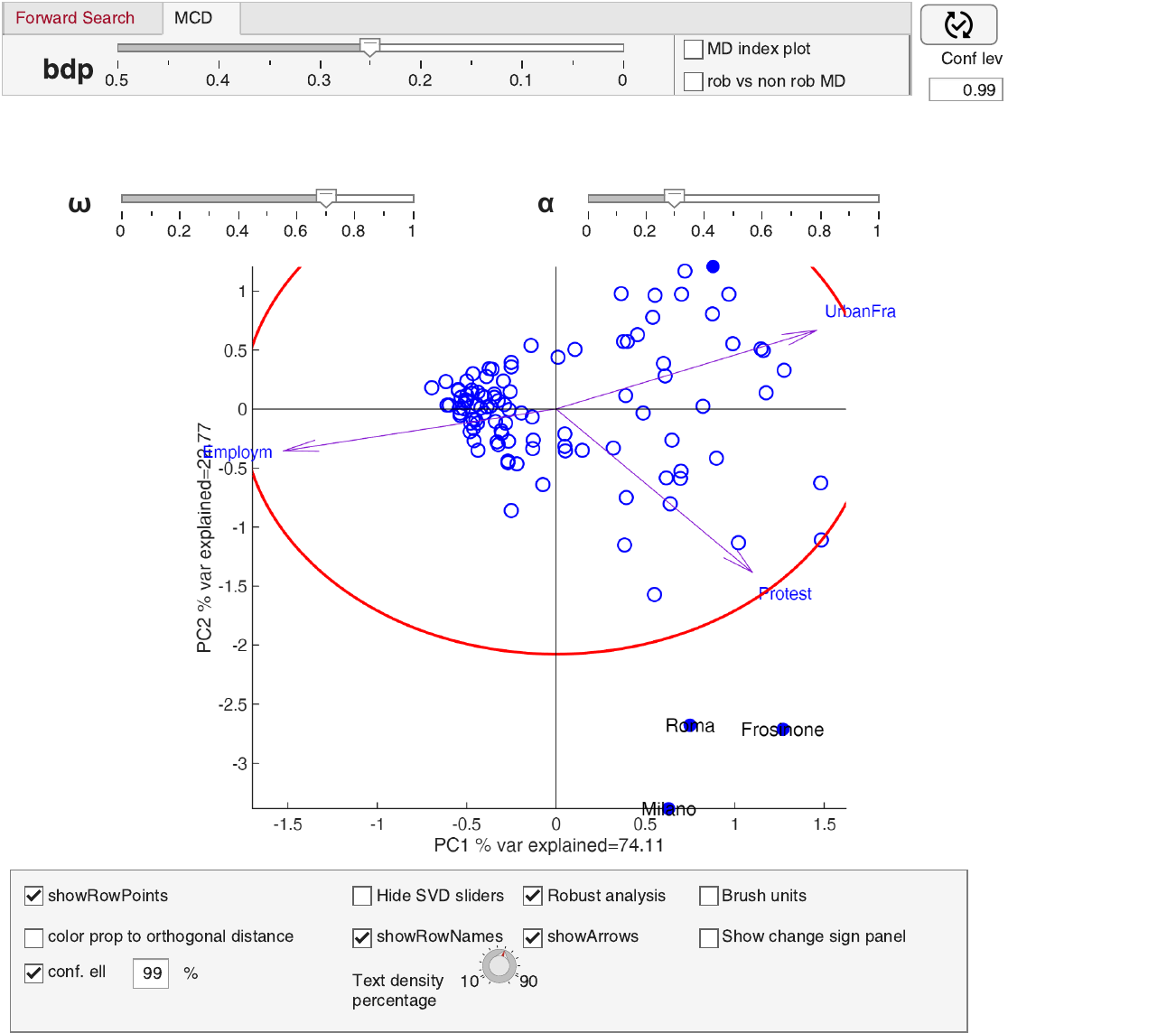}
\caption{MCD. Biplot based on just 3 variables using the units not declared as outliers with bdp = 0.25 and a simultaneous confidence level of 99\%. The fourth unit declared as an outlier at the top of the plot is the town of Caltanissetta (Sicily).} \label{APP3mcd025}
\end{figure}

Note that Figure~\ref{APPorthRowNames} retains the legacy palette (\mcode{'colorBlindSafe'} set by default to \texttt{false}) and thus illustrates
the previous graphical default rather than the revised palette policy. Under that policy, shown in the right-hand panel of Figure~\ref{APPorth}, ordered magnitudes
such as orthogonal distances use a perceptually ordered sequential HCL-based scale, and grouped or brushed observations combine color with marker shape or fill so that
interpretation does not depend on hue alone, following \citet{zeileis2020colorspace} and \citet{Crameri2020}.

Figure~\ref{APP3mcd} showed the robust biplot based on just 3 variables using the MCD with bdp = 0.5. Figure~\ref{APP3mcd025} shows the results of the calculation when the  bdp = 0.25. There are just four units declared as outliers from this less robust choice of the trimming parameter. The level of the simultaneous confidence ellipse is again  99\%. Note that the MCD has found outliers in the three-dimensional space of the variables, not in the space of the first two PCs.

\begin{figure}[t!]
\centering
\includegraphics[trim=0cm 22cm 0cm 0cm, clip, width=0.7\textwidth]{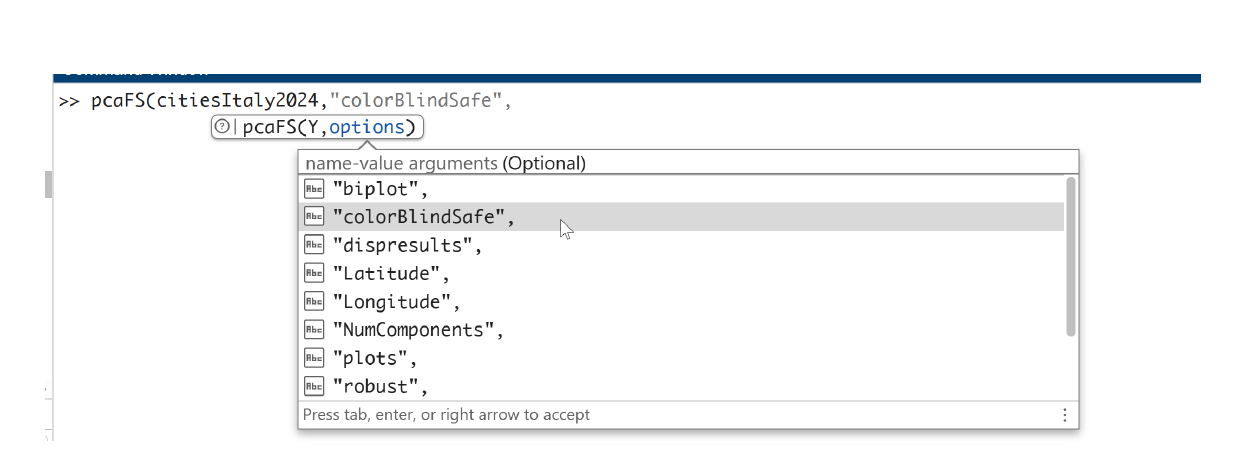}
\includegraphics[trim=0cm 22cm 0cm 0cm, clip, width=0.7\textwidth]{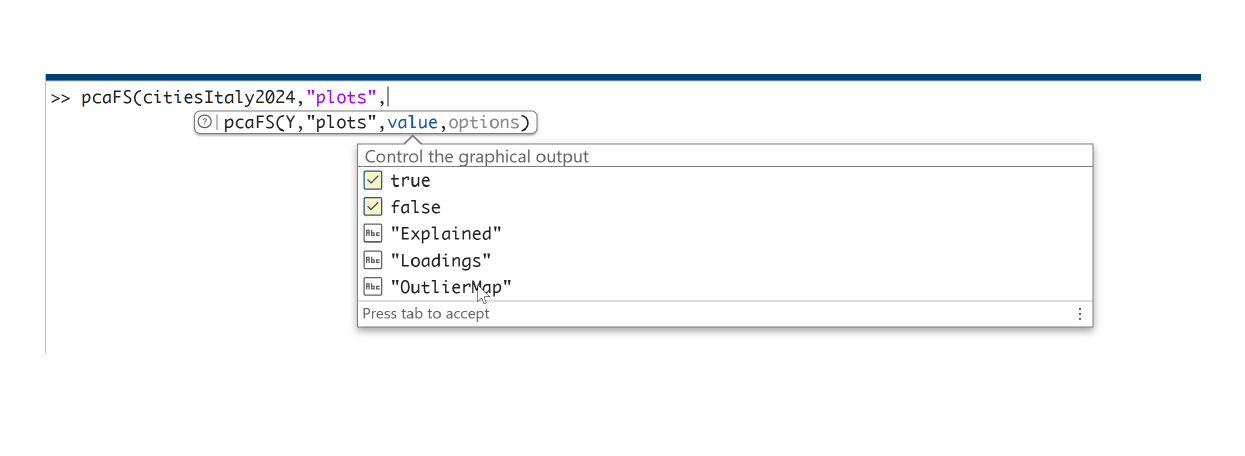}
\caption{Automatic completion of function options in the \proglang{MATLAB}
editor. The JSON signature file distributed with the FSDA implementation allows
the editor to suggest the available name-value arguments after the first
required input argument has been supplied (top panel). Once a particular option
has been selected, the editor suggests the admissible values or value types
(bottom panel). For example, the option \mcode{'plots'} can be specified either
as a logical value or as a string vector identifying the plots to be shown. This
makes the insertion of options easier and reduces the need to consult the help
file.}
\label{fig:jsoncompletion}
\end{figure}

\section{Output in the Command Window}
\label{sec:command_window}

In this appendix we present the output which appears in the Command Window (or inside the .mlx file) once the routine \texttt{pcaFS} is called.
For example, the call to \texttt{pcaFS} and the corresponding output are shown below:

%\begin{minipage}{\textwidth}
\begin{lstlisting}[firstnumber=1,basicstyle=\ttfamily\footnotesize]
Xsel=citiesItaly2024(:,{'Employm' 'Protest' 'UrbanFra'});
out=pcaFS(Xsel)
\end{lstlisting}
% \end{minipage}\\
% \begin{minipage}{\textwidth}
\begin{lstlisting}[basicstyle=\ttfamily\footnotesize , columns=fixed, commentstyle=\color{black}, stringstyle=\lstbasicfont\color{black}, numbers=none]
Initial correlation matrix
                Employm    Protest    UrbanFra
                _______    _______    ________

    Employm       1.00      -0.31      -0.90
    Protest      -0.31       1.00       0.23
    UrbanFra     -0.90       0.23       1.00

Explained variance by PCs
           Eigenvalues    Explained_Variance    Explained_Variance_cum
           ___________    __________________    ______________________

    PC1       2.04              67.98                    67.98
    PC2       0.86              28.75                    96.73
    PC3       0.10               3.27                   100.00

Loadings = correlations between variables and PCs
                 PC1      PC2
                _____    _____

    Employm     -0.96    -0.18
    Protest      0.49    -0.87
    UrbanFra     0.94     0.28

Communalities
                PC1     PC2     PC1-PC2
                ____    ____    _______

    Employm     0.92    0.03     0.95
    Protest     0.24    0.75     1.00
    UrbanFra    0.88    0.08     0.95

Combined OD-SD confidence level: 99.9907%
Squared combined OD-SD threshold: 18.556
Units outside the combined OD-SD confidence contour
                             Combined OD-SD
                             ______________

    Milano                         40.6
    Frosinone                    29.018
    Aosta                        28.593
    Roma                         26.156
    Barletta-Andria-Trani        21.787
    Taranto                      19.354

out =  struct with fields:

            Rtable: [3x3 table]
         explained: [3x3 double]
        explainedT: [3x3 table]
             coeff: [3x2 double]
            coeffT: [3x2 table]
          loadings: [3x2 double]
         loadingsT: [3x2 table]
     communalities: [3x3 double]
    communalitiesT: [3x3 table]
             score: [107x2 double]
            scoreT: [107x2 table]
          orthDist: [107x1 double]
         scoreDist: [107x1 double]
             class: "pcaFS"
\end{lstlisting}

\section{Overview of functionalities}
\label{sec:overview_functions}
\renewcommand{\thetable}{C.\arabic{table}}
\renewcommand{\theHtable}{C.\arabic{table}}
\setcounter{table}{0}
Table~\ref{tab:FSDAfunctions} complements Table~\ref{tab:functionality_main} by giving documentation links for the \pkg{FSDA} functions and interactive tools discussed in the paper. The main text describes the implementation design; this appendix is a compact reference for users who want to locate the corresponding help pages.
\begin{table}[h!]
\centering
\caption{Overview of FSDA functions discussed in the paper, with links to  documentation}
\label{tab:FSDAfunctions}
\vspace{3mm}
\resizebox{\textwidth}{!}{%
\begin{tabular}{|L{3.5cm}|L{5.5cm}|L{5.5cm}|}
\hline
\textbf{FSDA function / capability} & \textbf{Purpose} & \textbf{Illustrative use in paper} \\
\hline
\texttt{spmplot} &
Extended scatter plot matrix with variable ordering, correlation summaries, background colouring, and interactive options. &
Exploratory analysis of pairwise relationships and correlation structure (Section~\ref{datasec}, Appendix~\ref{sec:further_figures}). \\
\hline
\multicolumn{3}{|c|}{\url{http://rosa.unipr.it/FSDA/spmplot.html}} \\
\hline \hline
\texttt{pcaFS} &
Core routine for classical PCA with integrated numerical output, graphical displays, interactivity, robustness options, and linked brushing. &
Principal component analysis, scree plots, smart eigenvector displays, outlier maps, biplots, robustness diagnostics, and geographical visualizations (Sections~\ref{datasec}, \ref{indivsec}, \ref{robsec}, \ref{brushsec}). \\
\hline
Robust PCA via MCD (within \texttt{pcaFS}) &
Hard-trimming robust PCA based on the Minimum Covariance Determinant with user-specified breakdown point. &
Identification of outliers and comparison of robust and non-robust PCA fits (Section~\ref{robsec}). \\
\hline
Forward Search PCA (within \texttt{pcaFS}) &
Hard-trimming approach that monitors the evolution of PCA structure as the size of the subset used for the fit increases. &
Assessment of stability of principal components and influence of individual observations (Section~\ref{robsec}). \\
\hline
Robustness diagnostics (integrated panels of \texttt{pcaFS}) &
Monitoring plots of Mahalanobis distances, robust vs non-robust distances, and forward plots linked to PCA representations. &
Exploration of outlyingness and robustness as an exploratory dimension (Section~\ref{robsec}). \\
\hline
Geographical and brushing tools ('\texttt{ShapeFile}' option of \texttt{pcaFS}) &
Linked brushing between PCA plots, scatter plot matrices, geomaps, and geobubble representations. &
Spatial interpretation of PCA scores and robust structures (Section~\ref{brushsec}). \\
\hline
\multicolumn{3}{|c|}{\url{http://rosa.unipr.it/FSDA/pcaFS.html}} \\
\hline \hline
\texttt{biplotFS} &
Standalone interactive biplot with configurable scaling parameters, sign changes, confidence regions, and colouring options. &
Focused exploration of biplot representations, including reduced-variable analyses (Section~\ref{indivsec}). \\
\hline
\multicolumn{3}{|c|}{\url{http://rosa.unipr.it/FSDA/biplotFS.html}} \\
\hline
\hline
\texttt{pcaProjection} &
Three-dimensional visualization of observations, projections, and confidence ellipsoids in original and PCA spaces. &
Assessment of representation quality beyond two principal components (Section~\ref{3sec}). \\
\hline
\multicolumn{3}{|c|}{\url{http://rosa.unipr.it/FSDA/pcaProjection.html}} \\
\hline
\end{tabular}
}
\end{table}

\section{Compatibility with Octave}

%The core contribution of the present work lies in the development of tightly integrated interactive and dynamic graphical tools, including brushing, linking across multiple graphical views (biplots, scatter plot matrices, and geographical maps), real-time GUI controls, and event-driven callbacks. These features rely on MATLAB-specific graphics handles, UI components, and interaction mechanisms that currently have no functional equivalent in GNU Octave. As a consequence, full Octave compatibility for the interactive components is, at present, not technically feasible. That said, we note that the underlying computational routines for PCA are largely independent of the interactive layer. In particular, function \texttt{pcaProject}, which performs the numerical projection and transformation steps, has been made compatible with Octave if the routine is called using a 2D array.

The underlying computational routines for PCA in our work exhibit a significant degree of compatibility with GNU Octave. Notably, the function \texttt{pcaProjection}, responsible for executing the numerical projection and transformation steps, can function well within Octave when it is called using a 2D array, producing all informative plots. This allows users to employ the essential computational aspects of our developed tools even if they are using Octave.
However, there are limitations concerning the interactive components of our software. The core contribution of the present work involves the creation of integrated and dynamic graphical tools that feature capabilities such as brushing, linking across multiple graphical views (biplots, scatter plot matrices, and geographical maps), real-time GUI controls, and event-driven callbacks. These features depend heavily on MATLAB-specific graphics handles, UI components, and interaction mechanisms, which currently do not have equivalent functionality in GNU Octave. As such, full compatibility with Octave for these interactive elements is, at this point, technically infeasible.
We recognize these limitations but remain optimistic that future developments in GNU Octave may provide solutions that will allow for enhanced compatibility, overcoming some of the current obstacles in mirroring MATLAB's interactive graphical features.

\end{document}